\documentclass[10pt]{article}
\usepackage[ansinew]{inputenc}
\usepackage{fancyhdr}

\usepackage{amssymb}
\usepackage{tipa}
\usepackage[ansinew]{inputenc}
\usepackage{fancyhdr}
\usepackage[all]{xy}
\usepackage{sectsty} 
\usepackage[normalem]{ulem}
\allsectionsfont{\large}
\subsectionfont{\normalsize}
\subsubsectionfont{\normalsize}
\usepackage{amsmath}
\usepackage{mathenv}
\usepackage{dsfont}

\usepackage{graphicx}
\usepackage[colorlinks=true, citecolor=blue, linkcolor=blue, urlcolor=blue]{hyperref}
\usepackage[active,new,noold,marker]{xrcs}
\usepackage[active]{srcltx} 
\usepackage{color} 		

\title{\begin{flushright} \vspace{-70pt}\textit{\footnotesize Computer Science, Engineering \& \\ \vspace{-10pt} Info. Sciences, Phase: Theor. Proj.} \\
\textit{}\\\vspace{-20pt}
\textit{\footnotesize Prepr. Rep. on Comp. Sci.}$\,$\footnotesize \emph{\textbf{Ver.}} 2, 1--44\\ \vspace{2pt} \footnotesize Report Revised and Published on 14 Oct 2009
\\ --------------------------------------------------------------- \end{flushright} \vspace{12pt} \Large  \textbf{A Lossless Fuzzy Binary AND/OR Compressor} \\
}

\begin{document}


\definecolor{MyPurple}{rgb}{0.4,0.08,0.45}

\definecolor{MyBrown}{rgb}{0.6,0.4,0}

\definecolor{MyGreen}{rgb}{0.3,0.8,0}

\date{}
\maketitle
\begin{center}
\vspace{-50pt}
\long\def\symbolfootnote[#1]#2{\begingroup%
\def\thefootnote{\fnsymbol{footnote}}\footnote[#1]{#2}\endgroup}
\smallskip
By Philip Baback Alipour \noindent\symbolfootnote[1]{Author for correspondence (\htmladdnormallink {\textcolor{blue}{philipbaback\_orbsix@msn.com}}{mailto:philipbaback_orbsix@msn.com}).} $^{,}$ \noindent\symbolfootnote[2]{The current version is a preprint of the FBAR algorithm. The package development phase on very large databases including comparable results to this report, is aimed for future reports.
\vspace{-10pt} \begin{flushright} \textnormal \TeX Paper  \end{flushright}  \vspace{-28pt}}
\end{center}

\begin{center}
\textit{Department of Software Engineering and Computer Science, Research Project on Computational Engineering, Coding and Information Theory,}

\textit{School of Engineering Soft Center, Blekinge Institute of Technology,}\\
\textit{SE-372 25 Ronneby, Sweden}
\smallskip

\end{center}

\noindent \textbf{Abstract ----- }\small {In this report, a new fuzzy 2\emph{bit}-AND parallel-to-OR, or simply, a fuzzy binary AND/OR (FBAR) text data compression model as an algorithm is suggested for bettering \emph{spatial locality} limits on nodes during database transactions. The current model incorporates a four-layer application technique: \emph{string}-to-AND/OR \emph{pairwise binary bit} + \emph{fuzzy quantum with noise} conversions. This technique promotes a \emph{lossless data compression ratio} of 2:1 up to values $\thickapprox$ 3:1, generating a spatially-efficient compressed data file compared to nowadays data compressors. Data decompression/specific data reconstruction initiates an AND/OR pattern match technique in respect of fuzzy quantum indicators in the binary function field. The reconstruction of data occurs in the 4th layer using encryption methods. It is hypothesized that significant data compression ratio of $2n$:1 for $n$$>$3$:$1 ratios, e.g., 32$\sim$64$:$1 are achievable via \emph{fuzzy qubit indexing over classical byte blocks for every bit position fragmented into a $(\frac{1}{2} upper +\frac{1}{2} lower)$-bit noise frequency} parallel to its counterpart signal comprised of AND/ORed-\emph{bit polarity orientation}, ready for an identical data decompression.

\begin{center}
\vspace{6pt}
\smallskip
\end{center}
\vspace{-20pt}
\section{Introduction}
\label{section1}
\vspace{1pt}
\markboth{P. B. Alipour}{FBAR Compression Model}

\normalsize This paper addresses the fundamental information-theoretic data compression-decompression in locality of reference with entropic limits in terms of: \vspace{-2mm} \\

 \emph{Without losing 1-bit of information, is the problem of compressing data into a smaller space with short time representation of the same difficulty is as same as the problem of decompressing encoded data from a short representation?}   \\

\vspace{-2mm}

\noindent To have an efficient \emph{lossless data compression algorithm} implemented at a computer level, one must conceive the importance of cryptographic methods in aim of attaining certain efficient levels of data compression. These levels are given as application's data abstraction levels from source to sink, where sink is strongly time $t$-dependant and could be defined as the point of resource reconstruction, or, the point of sharing from source, or, a data bound by an external data such as a database (DB) that is made available by source, described by Patel\emph{ et al.}~\cite{22-Patel et al.}. Data abstraction, by definition, relates to `DB management system' with a complexity of physical, logical and file-view application levels such as size, 1-\emph{bit attribute flag}, \emph{binary search enumeration} (here, more of a nebulous order type) and \emph{bitrate} in DB transactions, via an installable data compressor at master/slave nodes interface, pp.$\,$4-6 of~\cite{22-Patel et al.}.

We shall, however, introduce our application on a physical level (\emph{binary} and \emph{fuzzy quantum}: let this computational composition of fields of logic be known as \emph{fqubit}) below a trivial monolayer of ASCII delivery level, in total, establishing a four-layer application. The layers in an array system of DB transactions could be defined as \emph{cache layering}, a concept well-recognized by Viana \emph{et al.}~\cite{31-Benevenuto}.

We formulate our algorithm by combining \emph{fuzzy logic} and \emph{quantum protocol} with noise characteristics which implies to every realistic communication system, p.$\,$165, Ref.~\cite{01-Joiner}. The prepended logic is crucially Boolean algebraic and in particular, we apply classical AND/OR logic for highly-convenient data compression results. In the current algorithm, we estimate to implement higher compression ratios \emph{without using lossy algorithms} up to approximately, at least 3$:$1 \emph{lossless compression}. This contrasts with the findings of Smith~\cite{14-Smith} which says: ``We have implemented a 3$:$1 compression ratio using a \emph{lossy algorithm}''.

Further to FBAR data compression characteristics, we also discussed the limits of \emph{fqubit} logic as FBAR's final product for highest data compression ratios beyond the ratios of current data compressors. It is at this level, FBAR becomes FQAR (\emph{fuzzy qubinary} AND/OR) due to quantum noise inclusions above FBAR \emph{fuzzy binary} classical layers. To this account, we have classified FBAR for \emph{classical computers} and FQAR for \emph{quantum computers}. Its proof of technology elicits from a combination of L. Zadeh's fuzzy set theory~\cite{{08-Zadeh,{09-Zadeh},{10-Zadeh}}}, \emph{qubit}, and \emph{classical binary} theorems. Henceforth, the term ``binary" should be understood as a sequence of ``classical bits" of information in particular. The use of \emph{binary} in virtue of \emph{qubit} becomes \emph{qubinary}, and is in contrast with the findings of Czachor~\cite{32-Czachor} in virtue of Abrams and Lloyd report~\cite{33-Abrams}, defining ``the idea of \emph{quantum computation}" which ``rests on the observation that a binary number $i_0 . . . i_{n-1}$ can be represented by a vector (a \emph{qubinary number}) $|i_0\rangle . . . |i_{n-1_i}\rangle$ denoting an uncorrelated state of $n$-distinguishable two-level quantum systems."

The notion of binary as qubinary in our model is due to its structural transformation FBAR$\rightarrow$FQAR, as a 1-\emph{bit vector} state with \emph{n}-dimensional binary sequence fragmentation in some \emph{qubinary} \emph{wave model representation} given in \S\S\,\ref{section3.4} and \ref{section4.1}. However, the qubinary early models involved nonlinear quantum algorithms for a quantum computer and have been discussed by Czachor in~\cite{32-Czachor}.

In this paper, we focus more on the classical approach and thereby aim for the quantum approach, herein the introduction and subsequently the FQAR in a future report parallel to this one. We contemplate FQAR physical products representing data as a qualitative form of FBAR quantitative degrees of promotion of binary conversions-to-compression addressing the structural transformation FBAR$\rightarrow$FQAR from one compression layer to another. The quality is obviously revealed for data integrity delivered at the data decompression stage. Thus, there is some room available for furthering FBAR compressions up to $2n$$:$1 ratios e.g., 32$\sim$64$:$1 lossless compression when quantum registers are redesigned and refined in context by a proposal; the basic details are given in \S\S\,\ref{section2}, \ref{section3.4} and \ref{section3.5}, formulating another report parallel to FBAR. The compatibility of $2n$:$1$ in context is mainly from FQAR and thus e.g., for $64$:$1$ denoting a 64MB file compressed to 1MB in a 1-\emph{bit frequency} made up of 2-\emph{half bits}, gives a
\[ {\rm Space\,Saving} = 1 - \frac{\rm Compressed\,Size}{\rm Uncompressed\,Size} = 1-\frac{8,388,608\, bits}{536,870,912\, bits}= 0.984375 \ {\rm or} \ 98\% \]

\noindent and with extreme compressions for large file sizes, the algorithm promotes such compression ratios to generate compression product degrees of $\approx 0 \%$. We have illustrated the ratio applicability in form of strings measured in fixed length of multiple bytes e.g., ``\verb"Philip"'' gives 6-\emph{bytes} = 48-\emph{bits}, a ratio of $48$:$1$ for FQAR final layers of the 4th layer reserved.

Finally, we proclaim the newly-developed algorithm despite of its infancy, at its prototypic level represents a novel study and highly competitive with newcomer compression algorithms using this combinatorial fields of \emph{logic}, integrated with \emph{quantum cryptography}, \emph{fuzzy logic} and \emph{classical binary}, inclusively.

\section{Main technique}
\label{section2}

During the conceptual investigation made upon the quantum aspect of the FBAR algorithm, the significance of Heisenberg's uncertainty principle (HUP) between crisp binary or boolean logic states came to our attention. By solely using conversion and Boolean logic operators, observable bound deviation of an initial data to its final form of compression, after a pregnantly imaginary time length $t=t+1-1=t'$ multiplied by the \emph{sine} of a space curve angle, engaged \emph{fuzzy and quantum behavior} for the FBAR model (between one encoded layer mapped to another). The $-1$ added to $+1$, gave us a \emph{bi-time intersection} to the behavior of gathering information on ongoing compressed data from one previous layer at ($t-1$) to its next, at ($t+1$), recalling any conversion or any encoded method used in the language of time translation. (See Lorentz \& Galilean~\cite{49-Poincare})

To successfully leave the FBAR system for good after the given time duration, as von Neumann's \emph{mean time} in Hilbert spaces for decompression~\cite{50-Reed}, one must reach data reconstruction between DBs. To satisfy this, quantum noise inclusions for co-occurring covariant signals~\cite{47-Debnath}, carrying compressed bits between DBs must be enabled before reaching any decompression. The carried bits are of fractional type, depending on how we design our quantum system with relevant components serving an \emph{infinite-dimensional separable spaces}, for all fragmented bits living in such spaces. The answer to this setup is conceiving the very concept of Hilbert space norm.

The Hilbert space is quite definitive in the satiated states' mappings of \emph{fractional occupance per bit} prior to \emph{byte} measurement (compare this to Smith~\cite{14-Smith}). That is, the unitary transformation of data complex or data index from one memory address to another. Especially, a growing space (\emph{variant}) reaching threshold points of absolute 0 and 1, their combination forms a definite Hilbert space $\mathcal{H}$, when a DB transaction occurs from source to sink in need of data partitioning, compaction and compression for efficient data local control, access and manageability. Data relativities of an \emph{n}-dimensional space equations and thus, quantum noise could be benefited in form of co-products of fuzzy quantum states, linking the pairwise binary sets to fuzzy and quantum sets. To conceive this concept, for now, let this HUP behavior to be regarded into the visual aspect of superimposition of \emph{n}-dimensional images (like \emph{interference pattern}) and not `matter' itself which is expressed in some superposition phenomenon~\cite{42-Dirac}.

\smallskip
\vspace{43mm}
\begin{flushleft} \hspace{-1mm}
\includegraphics[width=17mm, viewport= 0 0 10 10]{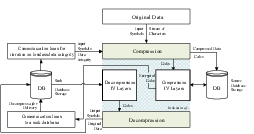}\\
\end{flushleft}
\vspace{-2mm}
\noindent{\footnotesize{\textbf{Fig.$\,$1.} This diagram is an expanded version of a transparent compression model presented in Fig.$\,$1 by Welch~\cite{07-Welch}. We adopted our algorithm to add the concept of fuzzy logic and Boolean logic with relative \emph{quantum noise} inclusions for the compression layers, thereby exclusions for the decompression layers. The output at decompression is equal to original data, later noted as $C'=C_{-1}$. The indication of `\emph{in domain of}' denotes that layers belong to either decompression or compression interface on the database storage system. To verify data integrity between source and sink databases, we further iterate the operation by revisiting the FBAR algorithm through lines of communications. This allows absolute entropic analysis for $2n$-bit/character I/O lossless data, preserved on nodes, both, quantitatively and qualitatively. }
\bigskip

\normalsize
In \S\,\ref{section4.1}, we further demonstrate ratio achievements of $2n$:1 for $n > 3$ through asymptotic behavior achievable via \emph{filename quantum indexing over byte lengths} when zero-byte file database managed through FBAR file/directory matrix configuration and allocation. \\

\vspace{4mm}
\begin{flushleft} \hspace{-1mm}
\includegraphics[width=10.3mm, viewport= 0 0 10 10]{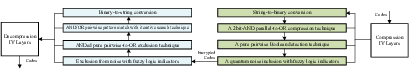}\\
\end{flushleft}

\noindent{\footnotesize{\textbf{Fig.$\,$2.} This image represents the dissected algorithmic components' decompression and compression sets of Fig.$\,$1. The flow is top-to-bottom on the compression's side and after encrypting the code, continues from bottom-to-top on the decompression's side. The expectation is, of course, outlined just like Fig.$\,$1, i.e. outputting original data in terms of $C'=C_{-1}$.}
\smallskip\\

\normalsize
This means, for 1TB data of any type is compressed by starting with 512GB and then $\approx$ 680GB up to $\approx$ 1TB lossless compression during probabilistic \emph{ergodic process}~\cite{{17-Shannon,{18-Shannon},{19-Birkhoff},{20-Anosov}}} for fuzzy quantum text distribution with a basic range of ]3:1, $2n$:1] estimate.} The intersection of space and string-to-binary format (\emph{a sequence of paired units of byte}) in programming, shall exhibit a \emph{dimensional finite system of compressed sequence} subject to the following: Given a Hilbert space $\mathcal{H}$, let for all $x$ pertain its \emph{reducible length function} $\lambda$ to the subsequent bilinear map
\vspace{-1mm}
\[ \forall \lambda \in \mathbb{C} \ \mathrm{and } \ \forall x \in \mathcal{H}^\textbf{*} \, ; \, (\langle\lambda(x)|)(|\lambda(x)\rangle) \in \mathbb{C} \ ,  \, \mathrm{where} \ \mathcal{H}^\textbf{*}\times \mathcal{H} \longrightarrow \mathbb{C} \, ,\]

\noindent by compression/decompression activities in the binary-fuzzy-quantum world of bit transformation, comes this

\[C(x) \sim C(y) \in \mathcal{H}\longrightarrow\mathcal{H}^\textbf{*} \sim C'(y) \in \mathcal{H}^\textbf{*}\times\mathcal{H}\longrightarrow\mathbb{C} \]

\noindent where projective values for data compression magnitude on $x$ in an encodable subspace of $\mathcal{H}$ as $\mathcal{H}^\textbf{*}$, for $x=2$ binary states elicited from the notion of Binary Logarithm~\cite{{28-Taylor,{29-Weisstein}}}, appears as
\vspace{-2mm}
\begin{equation}\label{1}
 C\left( x \right) = \parallel {x_{in} } \parallel \stackrel{t}{\to}  \parallel{y_{out} }\parallel  = \sum\limits_{i=1}^n \sqrt{\log_2 x_{i}^2} \stackrel{t}{\to} \mathop  \wedge \limits_{j = 1}^m \mathop  \vee \limits_{j = 1}^m \sqrt {\sum {\left( {y_j } \right)^2 } }\
\end{equation}\vspace{-4mm}

\noindent such that $\forall n\in\mathbb{Z}^+; \, n\geq 2m\, ,$ where $m$ is a number dedicated to AND/OR `$\wedge$ $\vee$' operators, and $t$ is the time needed to transform binary values from one compression layer to another. The notion of time, here, must not be confused with \emph{bitrate} which is a measurement of the number of \emph{bits} processed per unit of time. It is merely to compute \emph{entropy rate} and logarithmic behavior for its \emph{efficiency}. Thus, the AND-OR's \emph{bit population} representing 1 state resultant as a \emph{binary vector} being transformed from one compression layer to another, gives out \vspace{-4mm}
\[ C\left( x \right) = \left(1_1 + 1_2 + \ldots + 1_n\right) \stackrel{t}{\to} \left(\begin{array}{cc} 1_1 + 1_2 + \ldots + 1_{\frac{n}{2}} & 1_1 + 1_2 + \ldots + 1_{\frac{n}{2}} \\
\frac{1_1 + 1_2 + \ldots + 1_\frac{n}{2}}{2}   & \frac{1_1 + 1_2 + \ldots + 1_\frac{n}{2}}{2}  \\
1_{\frac{n}{\{n, n-1, ..., 5, 4  \}}} \pm \ldots   & 1_{\frac{n}{\{n, n-1, ..., 5, 4  \}}} \pm \ldots  \\ \end{array}\right) \]

\noindent whereas its time length function $\lambda(t)$ performing data compression between layers is partitioned in terms of
\begin{equation}\label{2}
C(x)=\frac{\lambda(x)}{\lambda(t)} = \left(\begin{array}{cc}
\frac{y}{t_1} & \frac{y}{t_1} \\
\frac{y}{2t_2} & \frac{y}{2t_2} \\
\frac{y}{4, 5, ..., n t_3} & \frac{y}{4, 5, ..., n t_3} \\
\end{array}
\right) = \lambda(y) \ , \, \mathrm{where} \ t_1 > t_2 > t_3 \ ,
\end{equation}

\noindent The bottom row of the matrix denotes a further or even an extreme compression after AND/OR application in the 4th layer satisfying conditions of $m<\frac{n}{2}$. The bottom left is not necessarily equal to bottom right in population, hence the use of $\pm$ notation. The left column indicates products of AND, and the right column indicates products of OR. The top left row is for AND, its neighboring right for OR from 2nd to 3rd layer compression. The latter layer satisfies conditions of $m=\frac{n}{2}$, whereas $n$ and $m$ in this layer act as a a positive integer; no matter the point of division on $n$, the value of $m$ in result is of positive integers. The middle row satisfies condition of \emph{pure pairwise detection of bits} preserving bit index $m=\frac{n}{2}$ whilst its binary equivalent \emph{string length function} $\lambda (x)= y = \frac{\emph{\texttt{len}}(\emph{\texttt{str}})}{2}$ from 3rd to 4th layer compression.

For data decompression magnitude, the binary vector transforming from one compressed or encoded binary result to one final decompression layer, retrieving/reconstructing original data on all $x$'s encoded product $f(x) \geq 2$ binary states ($x >2$ binary states, denotes fuzzy and quantum resultants after compression), would be
 \vspace{-2mm}\begin{equation}\label{3}
 C'\left( y \right) = \parallel {y_{in} } \parallel \to  \parallel{z_{out} }\parallel =\left(
                                                                                           \begin{array}{cc}
                                                                                            \frac{\lambda(y)}{2} & \frac{\lambda(y)}{2} \\
                                                                                             \lambda(y) & \lambda(y) \\
                                                                                             (2,\ldots , \mathbb{F}) \lambda(y) & (2, \ldots , \mathbb{F})\lambda(y) \\
                                                                                           \end{array}
                                                                                         \right)
 \end{equation}

\noindent The latter frame emphasizes the probabilistic values of inner-product of bit population residing in $x$ or $y$, or even both. The importance of our current product, from an $x$ to $y$ and thereby $z$, is in its vector space field, being either of real numbers $\mathbb{R}$ or the field of complex numbers $\mathbb{C}$, denoted by the field of scalars $\mathbb{F}$. This is due to some maximum point of division resulted into fragments of bit later known as \emph{data dots} in the quantum space. To this account, we could use the concept of \emph{affine subspace} or \emph{affine transformation} to express mappable binary data from a memory matrix (array). This data is quite connected from one conversion layer to another in a 2D plane-to-2Dplane geometric transformation in discrete time steps. This forms a 4D integration satisfying a lossless data decompression on $x$'s product in terms of $z$ sequencing after an input of compressed $x$ which is $y$. That would be

\begin{equation}\label{4}
\frac{{C_{ - 1} }}{{C'}}\left| {\frac{{C \stackrel{t}{\longrightarrow} C}}{{C \stackrel{t'}{\longrightarrow} C'}}} \right.\left| {\frac{{C'}}{{C_{ - 1} }}} \right.=\frac{\parallel {x_{in} } \parallel}{*}\left| {\frac{*}{ \parallel {z_{out} } \parallel}} = 1  \right.
\end{equation}

\noindent Relation (\ref{4}), commences with the Hilbert space bilinear map $\mathcal{H}^\textbf{*}\times \mathcal{H} \longrightarrow \mathbb{C}$, and associates with the projective values of our algorithm from pairs of bits in the binary sequence once a conversion is made by either. This value projection onto the $xy$ plane is of compression $C(i,j)$ or decompression $C'(i,j)$. Let the plane of compression exhibiting integrable areas satisfy compressed indices $i$ and $j$ of matrix $A$, such that for each layer of the lossless compression algorithm abides by the rules of binary ranks over a field of binary functions. Once a layer is encoded after a duration of time $t$, the rank if, is of compression phase, it is said that more data from the binary sequence is compressed. If the rank is of decompression phase, the same data is decompressed without loss of data regardless of entropy 2 \emph{bit/character} application; or ratio 2 $ bit/(byte, nibble) = \{0.25 , 0.5\}$, where the nibble case implies to a space character due to having 4 pure paired 0's which could be compressed to 0000, later experienced in our algorithm. We initially for such rankings of a conversable (0,1)-matrix, studied the Rigidity Theorems of Hamada and Ohmori, revisited by Mickael, pp.$\,$175-179, Ref.~\cite{01-Joiner}. The expanded form of Rel.$\,$(\ref{4}) applying AND/OR logic, fuzzy and quantum definitions relevant to the flow presented in Fig.$\,$2, deduces $C' = C_{ - 1}$, otherwise is an unbalanced relation without a reliable encoded techniques applied. \\

\noindent \emph{Ranks of compression and decompression: }----------- The ranking characteristics are tangibly illustrated in Fig.$\,$2, establishing \textbf{0} as base of mappable\emph{ binary }and \emph{fuzzy} functions between Hilbert space dimensions; \textbf{1} represents fixed values of binary with their \emph{quantum noise} inclusions including pure pairs i.e. 00 and 11 mapped into 01 \emph{fuzzy state} during logic coding. This fuzzy state promotes 2 \emph{bits/character} to 4 \emph{bits/chacater} compression, a set $\{0.5, 0.75 \}$ \emph{space saving} is obtained with a fine line of symmetry between $\{00, 01, 11\}$ and $\{11, 10, 00\}$ out of a possible set of binary combinatorial states $\mathbb{B}^{01} = \{01, 00, 10, 11\} $. There is no repetition of a pure state in regard to the established fuzzy state 10 and 01, since the 1st ranked 11 is the mirror of the 2nd ranked 11, and the 1st ranked 00 the mirror of the 2nd ranked 00, gives out for $C'$, a 11 and 00 decompressed values, after lower bound binary conversions (layer $\ell'$ slopes after $t$-seconds). In fact, the 2nd ranked set could be delivered as decompressed data preserving all data content. To this account, in principle, we logically state for all bit combinations' base \textbf{1} to be in the 1st layer $\{\ell\}$ with 1st ranked as $C_{-1}$ or no-compression (original data \emph{before} $C$ function, hence the use of $-1$ in its index), and set of 2nd ranks as $C'$. Therefore, all base-bit sequences of $C_{-1}$, $C$ and $C'$, baseline \textbf{0} are in 2nd and 1st layers $\{\ell\ell'\}$ during all complementary $C_{-1}CC'$ conversions. Presentable values of key-line denoted by $\textbf{*}$ (one could also say, negative or imaginary numbers line for all $f\in \mathcal{H}^\textbf{*}$), are in purely 2nd-to-4th layers, or respectively  $\{\int\ell\ell\longrightarrow\ell''\mathrm{d}t\}$ during just compression $C$ conversions before $CC'$ initiations. The time factor is always $t + 1$ for these layers, and after $t$-seconds, reverting back into 1st ranked values of $C_{-1}$ by mapping quantum + fuzzy values of $C$ to $C'$. The objective is obviously obtaining $C'=C_{-1}$ in practice. Note this as a quantum exclusion, to obtain $C'$ lossless data at the point of sink, which once was as $C_{-1}$ at the point of source. Finally, the points of closure to all projective mappings from layer $\ell$ to $\ell'$ and $\ell'$ to $\ell$, so on so forth, are of $\ell''$, which is strictly entopic-related conserving the amount of data from $C$ to $C'$, and vice versa. The layer is spatially-timely connected to layers of compression, since the inclusion of quantum noise is excluded when decompression occurs for the binary sequence (observe Fig.$\,$3, baseline \textbf{0}). In addition, the string length function $\lambda (x\vee y \vee z) = \emph{\texttt{len}}(\emph{\texttt{str}})$ as a solution to binary values is given in a set of pairs $\lambda =\sqrt{4n^2}$ or $2n$ projected orthogonally in $\{\int\ell\ell\longrightarrow\ell''\mathrm{d}t\}$ and is displayed later in form of a quadruple integral. The binary values conversion $C$, resulting under the area of a compression curve, integrates into layer $\ell''$ of the dual Hilbert space $ \mathcal{H}^\textbf{*}$. The result in $C$ would then be in terms of pairs of increments $\Delta \lambda = \emph{\texttt{len}}_{--}(\emph{\texttt{str}})=\sqrt{4n^2 - m^2}$ such that, transformation $C\rightarrow C'$ gives $\Delta\Delta \lambda = \emph{\texttt{len}}_{(--,++)}(\emph{\texttt{str}})=\sqrt{(4n^2 - m^2)} \longrightarrow \sqrt{(4n^2 + m^2\imath^2)}$, where $\imath = \sqrt{-1}$.  -----------

\subsection{Zadeh operators revisited}
\label{section2.1}

We revisit Lotfi Zadeh's fuzzy logic. The term ``fuzzy logic" emerged as a consequence of the development of the theory of fuzzy sets by L. A. Zadeh~\cite{03-Fuzzy}.

In 1965, Zadeh proposed fuzzy set theory~\cite{10-Zadeh}, and later established fuzzy logic based on fuzzy sets. For example, an extremely simple temperature regulator that uses a fan might look like this:


\footnotesize{\begin{flushleft}
------------------------------------------------------------------------------------------------------------------------
    \texttt{IF temperature IS very cold THEN stop fan} \ \ \ \ \ \ \ \ \ \ \ \ \ \ \ \ \ \ \ \ \ \ \ \ \ \ \ \ \ \ \ \ \ \ \ \ \ \  \\
    \texttt{IF temperature IS cold THEN turn down fan} \ \ \ \ \ \ \ \ \ \ \ \ \ \ \ \ \ \ \ \ \ \ \ \ \ \ \ \ \ \ \ \ \ \ \ \ \ \  \\
    \texttt{IF temperature IS normal THEN maintain level }\ \ \ \ \ \ \ \ \ \ \ \ \ \ \ \ \ \ \ \ \ \ \ \ \ \ \ \ \ \ \ \ \ \ \  \\
    \texttt{IF temperature IS hot THEN speed up fan} \ \ \ \ \ \ \ \ \ \ \ \ \ \ \ \ \ \ \ \ \ \ \ \ \ \ \ \ \ \ \ \ \ \ \ \ \ \ \ \  \\
------------------------------------------------------------------------------------------------------------------------
\end{flushleft}

\normalsize
Notice there is no ``ELSE''. All of the rules are evaluated, because the temperature might be ``cold" and ``normal" at the same time to different degrees. We herein expanded the notion of the AND, OR, and NOT operators of Boolean logic in terms of AND, OR, AND/OR application over paired bits and the complement, quite evident in the layers of compression $C$ as described in the previous paragraphs prior to decompression $C'$ and original data $C_{0}$. The IF statements in our program should eventually appear as follows:

\footnotesize{\begin{flushleft}
------------------------------------------------------------------------------------------------------------------------
    $\texttt{IF logic IS 0 paired logic THEN DO AND 00}$\\
    $\texttt{IF logic IS 01 paired logic THEN DO AND OR}$\\
    $\texttt{IF logic IS 10 paired logic THEN maintain AND/OR level}$\\
    $\texttt{IF logic IS 1 paired logic THEN DO OR 11}$\\
------------------------------------------------------------------------------------------------------------------------
\end{flushleft}}

\normalsize The AND, OR, and NOT operators of Boolean logic exist in fuzzy logic, usually defined as the minimum, maximum, and complement; when they are defined this way, they are called the Zadeh operators, because they were first defined as such in Zadeh's original papers~\cite{08-Zadeh}. There are also other operators, more linguistic in nature, called hedges that can be applied. These are generally adverbs such as ``very'', or ``somewhat'', which modify the meaning of a set using a mathematical formula. \\

%
%
\noindent \emph{Further development:} Once fuzzy relations are defined, it is possible to develop fuzzy relational databases. The first fuzzy relational database, FRDB, appeared in M. Zemankova's dissertation. Later, some other models arose like the Buckles-Petry model, the Prade-Testemale Model, the Umano-Fukami model or the GEFRED model by J. M. Medina, M. A. Vila \emph{et al.} In the context of fuzzy databases, some fuzzy querying languages have been defined, highlighting the SQLf by P. Bosc \emph{et al.}~\cite{27-Bosc et al.}, and the FSQL by J. Galindo \emph{et al.} These languages define some structures in order to include fuzzy aspects in the SQL statements, like fuzzy conditions, fuzzy comparators, fuzzy constants, fuzzy constraints, fuzzy thresholds, linguistic labels and so on. Further development to the FBAR algorithm would aim to quantify queries on DBs in form fuzzy qubinary relational DBs. In fact, FBAR as FQAR would attempt to achieve highly efficient degrees of data reconstruction on a massive scale relative to multi-core architecture optimization.

\subsection{Model specification}
\label{section2.2}

The current model's chronological abstractions from one compression layer to another attaining a level of reliable decompression with respect to entropic characteristics are:

\begin{enumerate}
\item [{1-}] Adjacently applicable fuzzy state once Boolean logic implemented for a four-layer compression before a conversed compression (decompression) from an $n$-dimensional Hilbert space norm. Fuzzy state applicability is of pre-layering done after converting string of characters-to-binary at $t = t - 1$ seconds.
\item [{2-}]Converting the binary quantities into sub-binaries to balance the conversion criteria over all conversions of bits of binary into strings with a ratio of values aiming at $\approx0 $ bits for maximum compression, and vice versa, $>0$ bits, denoting the original data after its decompression state.
\item [{3-}]Delivering the right data of decompressed string values with better efficiency to the users of destination from source values into the same string values via communication lines.
\item [{4-}]Entropic behavior of binary finite planes of lossless compression ratio on prepended data in bit-rate (bit/second) measurements. For instance, the converted symbol ``P" to binary from the following section, 01010000, is replaced by 0010. The length of the bit sequence ($4_{8}=0010_{2}$) is prepended. We hypothetically computed the lossless compression ratio in the FBAR algorithm to be classically 2$\thicksim$3:1, and with quantum features, $2n$:$1$ much greater than the well-known data compressors on the market today.
\end{enumerate}

\section{Main technique's four-layer application}
\label{section3}

The new algorithm obeys a certain logic prefiguratively illustrated in Fig.$\,$3, outlining the concept from its very principles based on time and process of data including the main technique per se. This figure represents configurable binary dataset after trivial monolayer conversion of string-to-binary in some programming language. We later program the converting routine in C language (any other language is an alternative to prove the FBAR algorithm, here, VB).

In Fig.$\,$3, the meaning of the expressions Low state, Combinatorial High-Low state, and High state logic is represented by functions mapping a compression-decompression scale. A point on that scale has three ``truth values'' — one for each of the three functions. The vertical line in the image represents a particular compression-decompression conversive phase, the base of a specific state indicated by the three arrows (truth values) gauge. Since the descending arrow emerging from 1.0 points to levels of the asterisk zone, this state may be interpreted as ``purely zero''. The upper-left arrow (pointing to 0.2) may describe it as ``half-side zero'' and the top-left arrow (pointing to 0.8) ``fairly half-side zero'' whereas its half-side zero occurs after being concatenated with its other half either of High (forming \emph{binary pair} 01) or Low (forming \emph{binary pair} 10) otherwise purely zero through AND/OR application. The asterisk symbol, `` \textbf{*} ", represents 2nd to 4th layers including decompression layers from 4th to 2nd layers before decompression sequence. Lossless conversion is thereafter initiated for the 2nd thus 1st layers performing mostly 2nd ranked and least of 1st ranked decompressed binary values, in this case, 00 and 11, via 01 and 10 as possible fuzzy combination, altogether satisfying a $2^3$ possible logic states: three of 2nd rank or $\{00, 10, 11\}$, and one of 1st rank or $\{01\}$. The in-between values (real values) like the 0.2, 0.8 and ... are contemplated for quantum noise inclusions and exclusions on FBAR data mechanisms approaching and thereby defining the three states. Such conditions are necessary to establish in the program code during extreme levels of FBAR data compression mechanisms.

\smallskip
\vspace{64mm}
\begin{flushleft} \hspace{-1mm}
\includegraphics[width=1.78mm, viewport= 0 0 10 10]{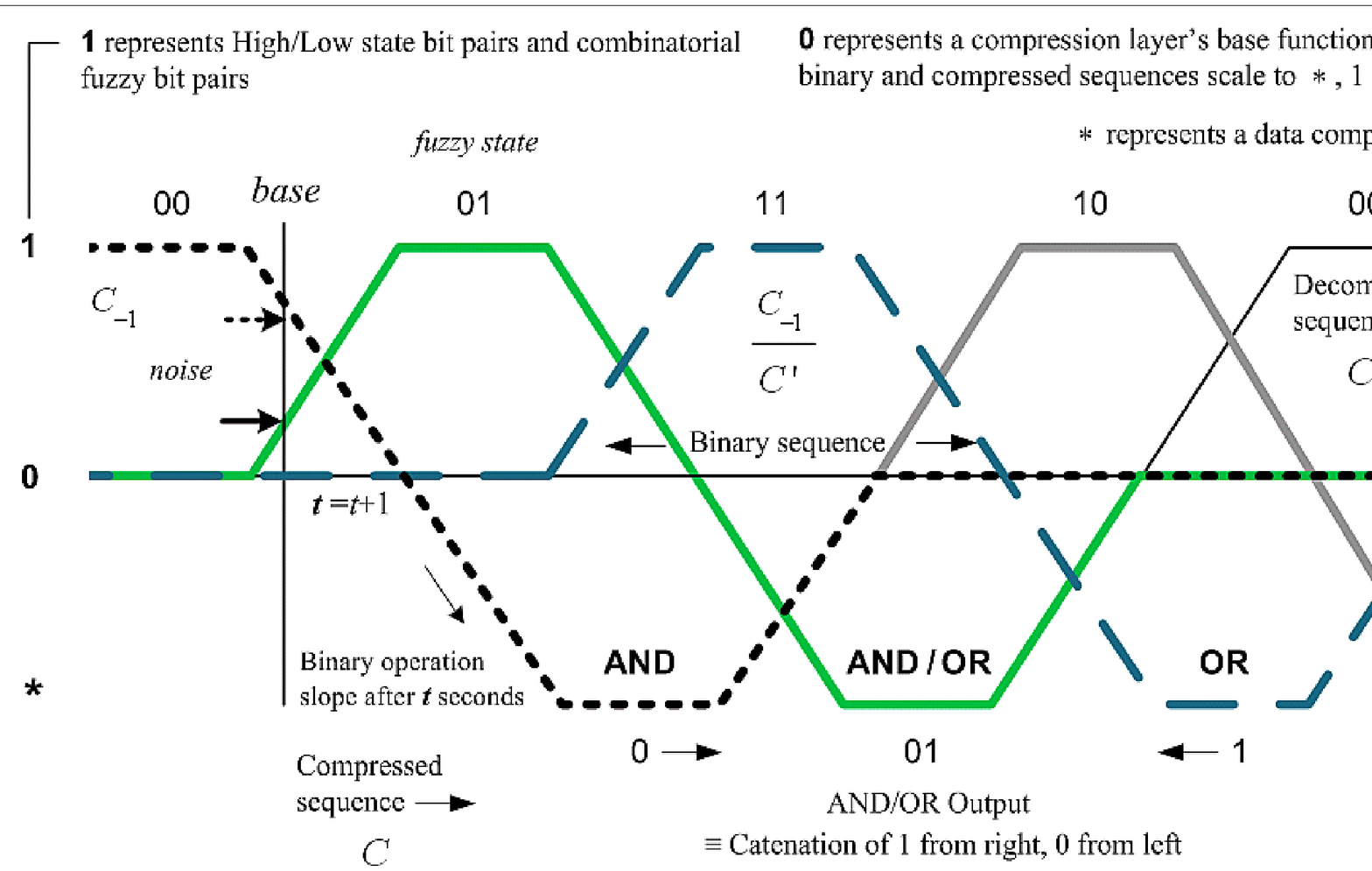}\\
\end{flushleft}
\vspace{-2mm}
\noindent{\footnotesize{\textbf{Fig.$\,$3.} A comprehensive fuzzy binary logic illustration displaying ``purely zero'', ``half-side zero'' and ``fairly half-side zero'' states (towards \emph{binary pairs} \texttt{01} or \texttt{10}, otherwise \texttt{0}) through AND/OR application. In-between states is considered quantum noise inclusions and exclusions on FBAR data mechanisms approach and thereby defining the three states.}
\bigskip

\normalsize
We begin by taking a string of ASCII characters, author's forename ``\verb"Philip"'', assuming indexed into a list of data atoms on a temporary database (TDB). The below quadruple integral expresses paired values of bit indices when converted from the string set
\[\Sigma^* =\bigcup\limits_{n\in\mathbb{N}}{} \Sigma^n ,\]

\noindent where a set of all strings over $\Sigma$ of any length is the Kleene closure of $\Sigma$ and is denoted $\Sigma^*$, given in terms of $\Sigma^n$. In this case, we convert the notion of string length measured by the number of characters in its set in terms of bits. Hence, we apply, for example, a length of $n=8$ bits to binary set $\Sigma^8$ for 6 characters generates $\{01010000, 01101000, 01101001, $ $ 01101100, 01101001, 01110000 \}$ compressing the set elements in form of pairwise combinations $\Sigma^2 = \{ 00, 01, 10, 11 \} $ between $xyz$ dimensions, treating the source dimension projecting bits of $\Sigma^8$, as a constant during pairwise transformation. Thus, $\forall x, y, z \in C_{-1},$ we then have
\begin{equation*}
\iiiint f(\ell, \ell', \ell'',t) \ \mathrm{d}\ell \,\mathrm{d}\ell'\,\mathrm{d}\ell'' \mathrm{d} t  = \ \  \ \ \ \ \ \ \ \ \ \ \ \ \ \ \ \ \ \ \ \ \ \ \ \ \ \ \ \ \ \ \ \ \ \  \ \ \ \ \ \ \ \ \ \ \ \ \ \ \ \ \ \ \
\end{equation*} \vspace{-10pt}
\begin{equation} \label{5}
    \mathop {\lim }\limits_{N \to \infty } \frac{1}{N} \sum_{n=0}^{m=N-1}  \sum_{m}^{q=\frac{n-1}{2}}  \sum_{q}^{r=\frac{q}{N\rightarrow\infty}} \sum_{t=0}^{t} \frac {U^{n} C_{-1}}{U_{t}} \delta \ell \, \delta \ell'\, \delta \ell''\, \delta t = \frac{P^2 C_{-1}}{2P}  = 0   \ \ \ \ \ \ \ \ \ \ \
\end{equation}

\noindent When $n$ approaches infinity in the \emph{mean ergodic theorem}~\cite{21-Reed}, the compression $C$ limit in integration aims at string length values $\lambda\geqslant 0$ for complete data sequence average of $n$ with respect to time $t$ giving a perfect data compressor impending  the total integral progression between layers. $P$ is an orthogonal projection between probable compression layers. Therefore, $P^2$=$P$ is \emph{idempotent}, proving that $P$ within the context of bitrate projection per character is indeed a \emph{projection} for $f(x) \in \mathcal{H}^\textbf{*}$ when $\emph{\texttt{len}}(x)\rightarrow0$, $\emph{\texttt{len}}(y)=C(x)\rightarrow0$ and $\emph{\texttt{len}}(z)=C'(xy)$ respectively. Also, let $U_t$ be a strongly continuous one-parameter group of unitary operators on Hilbert space $\mathcal{H}$ and converges in the strong operator topology as $T\rightarrow\infty$. One must conceive that all results are in respect of the pigeonhole principle resulting a statistical \emph{entropy rate} measured in bits/sample for a 27-possible values in the English alphabet $\mathcal{A}$ letters (characters), $\mathcal{A} = \{a , b , c , d , e , f , g , h , i , j , k , l , m , n , o , p , q , r , s , t , u , v , w , x , y , z , \mathrm{null}\}$.

In the above quadruple integral, this theorem is adapted into the following entropy rate for an extremely maximum compression, $C_{\max}$, \emph{if and only if} a quantized version of all compressed bits into \emph{fqubits} signal criterion elicited and hereby expanded to the \emph{fqubit} model representation (Fig.$\,$6 on p.\,22). For 1-\emph{fqubit} signals relative to the probabilities of entropy of a continuous distribution, \S\S\,20, 21 and 22 of~\cite{18-Shannon}, assuming $x$ is limited into a certain volume $\mathbf{v}$ in its space, in our case ``\emph{quite atomic}" or vertex-like $v$, the Hilbert boundary is maximum for at least projecting its product values into its neighboring $y$ having identical specification. So all probabilities of $x$ as $\mathbf{p}(x)$ are a constant 1 (certain to occur) in the volume $v$ for their occupying values (later known as \emph{data dots}) each as ($1/v$) in the volume. If the same point of volumetric division applies to all neighboring functions with spatial density, let the same $y$ in the infinite space norm conjugate the same space for all ANDed and ORed values including their duals \emph{projected} from $x$ to $y$. Thus, we let a vector value of 1 filled or if $v_y$ = $v_x$ then $1/v_x/v_y$=1 for $y$'s physical space occupied by values of $x$ in our matrix. We then derive
\vspace{0mm}
\[
C_{\max}= \sum_i (P_\psi (xy))_i = \lambda\left(\psi\psi^\top \left( \begin{array}{cc}
                                                            x & x \\
                                                            y & y
                                                          \end{array} \right) \right)= \sum_{i} \frac{bit_i}{\prod_i 2_i}\left( \begin{array}{cc}
                                                            0 & 0 \\
                                                            1 & 1
                                                          \end{array} \right) \ \ \ \
 \ \ \ \ \  \]
 \vspace{-2mm}
 \[=  \left\{\{0,0\}, \left\{\max( C_{\wedge,\vee}), \max(C_{\alpha,p}) \right\} \right\} \ \ \ \ \ \ \ \ \ \ \ \ \ \ \ \ \ \ \ \ \ \ \ \ \ \ \ \ \ \ \ \ \ \ \ \ \ \]
 \noindent where $\max( C_{\wedge,\vee})$ for maximally compressed ANDed and ORed values relative to the data on their \emph{address} $\alpha$ and \emph{polarity} $p$ in terms of $\max( C_{\alpha,p})$ (these bit properties are subject to \S\,\ref{section3.4}), have a fragmented size of
\[\max( C_{\wedge,\vee}) = \frac{1_1}{2_1 2_2 2_3, \ldots, 2_n}+\frac{1_2}{2_1 2_2 2_3,\ldots ,2_n}+ \dots + \frac{1_{n}}{2_1 2_2 2_3,\ldots, 2_n} \approx 0, \]
\begin{equation} \therefore \label{6}\max(C_{\alpha,p}) \in \, ]0, 0.5]  \, . \ \ \ \ \ \ \ \ \ \ \ \ \ \ \ \ \ \ \ \ \ \ \ \ \ \ \ \ \ \ \ \ \ \ \ \ \ \ \ \ \ \ \ \ \ \ \ \ \   \ \ \ \ \ \ \ \ \ \ \ \ \ \ \ \ \ \ \ \ \ \
\end{equation}

\vspace{2mm}
\noindent where $C_{\max}$ for all summed or totally grouped $x$'s and $y$'s, is one kind of continuous distribution containing projection $P$ of a wave-function $\psi$ values parallel to a coupled $\psi$ distribution, orthogonal to a certain point of ultimate projection. This ``ultimate projection" satiates values of $x$ in a matrix row and $y$ in a column format, whereby \emph{orthogonal optical pumping} matters between $x$'s and $y$'s micro-potential degrees. The result would be an $xy$ product associated with all degrees of projection from $x$ to $y$ satisfying $\lambda(\psi(x,t))$ conversions.

The final projection onto the wave for micro-potentials is discussed in \S\,\ref{section3.4}, which implies to all of registered bits reaching values close to 0 onto the maximally compressed planes that justly occupy a dot. In other words, to register a dot, $\forall xy_{bit} \ll xy$, where $xy_{bit}$ denotes the logic state of a bit that has already reached 1 or 0 logic, e.g., $xy_{bit}=0 \equiv 0 \in \psi(x,t=0)$, and, $xy_{bit}=1 \equiv 1 \in \psi(x,t=0)$. For other forms representing unitary transformations between two $\psi$'s, with a selected clocked length $\lambda$ on each wave, or, $\lambda(\psi_i)$  we have, e.g., if $(xy_{bit_{1}}=0,xy_{bit_{2}}=0) \longmapsto  xy_{qbit}$ then $U: \lambda=  |\psi_0\rangle \longmapsto |\psi_1 \rangle \equiv(0,0) \in \psi((y_{21},y_{22}),t=1)$. The latter form is referenced to Eq.\,(\ref{5}).

Therefore, using Eq.\,(\ref{3}), in the ``\verb"Philip"'' example, 48 classical bits are used to represent 6 characters giving  $(((((((48/2)/2)/2)/2)/2)/2)/...)/2$ bits/character for a quantized signal representing all compressed bits of information at a binary fuzzy quantum level. Ergo, Eq.\,(\ref{3}) as a fixed joint of quantum values to this deportment between $xy$ planes is amended/simplified down to
\begin{equation}\label{7}
 \therefore C'\left( y \right) = \parallel {y_{in} } \parallel \to  \parallel{z_{out} }\parallel =\left(
                                                                                           \begin{array}{cc}
                                                                                            \frac{\lambda(y)}{2} & \frac{\lambda(y)}{2} \\
                                                                                             \lambda(y) & \lambda(y) \\
                                                                                             2\lambda(y) & 2\lambda(y) \\
                                                                                           \end{array}
                                                                                         \right)
  \equiv 2^x \frac{\lambda(x)}{2} = 2^{(x-1)} y \ ,
\end{equation}
\noindent  since the notion of scalars field $\mathbb{F}$ is by now supposedly affine  between $x$'s and $y$'s transformation per bit for all complex and real values upon the proof of an existing dot in space. The above relations corroborate with ``Entropy Rate of a Source'', and the logarithmic state fixates onto that: the higher the order, the lower the rate (better compression). We are only interested in lossless data compression code for the FBAR model. That is, given the code table resulted from binary OR/AND and fuzzy quantum inclusions, and given the compressed data, we should be able to rederive the original data or $C_{-1}$. All of the examples given above are lossless.

The `$=0$' condition from the FBAR mechanisms (Fig.\,3) denotes that a quantum noise could be included but not absolutely. The quantized inclusion of noise, formulates a conditional value of threshold voltage in logic gates indicating either 1, 0 or even both. Hence in the absolute ``$=0$" condition, we define a zero value condition, not 1, a total 0 in area under a signal curve. The reason is that Boolean 0 for a gate could be a ``low" value operating between 0 V and 0.8 V, and for Boolean 1 or ``high", between 2.2 V and 5 V with a +5 Volts power supply e.g., TTL (transistor-transistor logic) gates. Ergo, for a quantum gate, clocked signals are in-between states for every binary sequence containing more than two bits, \emph{a union between these ranges of Volts} converted to values in the two-level quantum mechanical system i.e. Bloch sphere. The 1-\emph{qubit} could in fact represent the minimally two bits' logic states either each carry 0 logic or 1 logic, the qubit carries both 0 and 1 simultaneously. In other words, given two natural numbers $n$ and $m$ with $n > m$, if $n$ items as bits in the binary sequence are put into $m$ pigeonholes as quantum registers, then at least, one pigeonhole (a register cell) must contain more than one item (\emph{at least two classical bits into one quantum bit}). The qubit register should at least represent more than 1-bit's logic i.e., minimally 2-logic states probability for that same bit becomes $2^3=8$ probability of logic states for each \emph{qubit}. However, for an \emph{fqubit}, a multi-leveled mapping system is plausible if and only if its minimal probability obeys first order derivative implication between two binary planes of type $\mathbb{B}^{01}$ or $\Sigma^{8}$ with $\Sigma^2$, as
\vspace{-2mm}
\begin{equation}\label{8}
    \mathrm{if}\ \delta_C=\frac{\Sigma^{f'\left(8n^{2}\right)}}{\Sigma^{f'(2)}}\equiv\frac{\log_{2}(16)}{\log_{2}(0)}= \frac{4}{-\infty}=0; \  \delta_C \stackrel{C'}{\longrightarrow} \parallel\lambda_i\parallel = \sqrt{\prod\limits_{i = 1}^2 {\frac{8n^{2}}{2}}} = 4n^{2}
\end{equation}
\vspace{-1mm}

\noindent which says that the algorithm at first corresponds to 2 bits/character and is based upon the quadruple integral Rel.\,(\ref{5}), then when projection applies, becomes a value of a dot in the projected bit, whereas this bit is now former in time ($t=0$) and equal to 0 whilst a dot, a true value of $>$ 0 at $t+1$. The notion of time $t$, is measured by layers $\ell, \ell', \ell''$ when processed from one conversion layer to another between data code phases (system steps) akin to the model on p.$\,$9,~\cite{07-Welch}. Compression and decompression products of the algorithm denote these $\log_2 (v,(1/v))$ code phases from the model, now adapted to FBAR algorithmic model. The total phases' adaptation resulting a product of $f(x) = x^2$, in terms of derivative $f'(x)=f(n)=\sqrt{4n^{2}}=2n$ maps values of complex plane via $m$ from p.\,7, and perfectly fits in the polynomial curves, performing symmetric values on interpolation of $C$ and $C'$ correspondingly.

Thus the notion of shaping a curve, like a quantized signal, becomes relevant in these interpolations, and once the weight of projected dots distributes across the curve (\emph{signal}), accurate interpolations are formed, specifically, with \emph{asymptotic density} (on the interval $[-1,1]$ between $\pi$ revolutions) given by $1/\sqrt{1-x^2}$, in Berrut \& Trefethen paper~\cite{48-Berrut}. In our case, the  function of either $n$ or $m$ is substituted for $x$, and the complex planes representing the off scale limits of distributed dots are on the superimposed interval $[\cos(3\pi/2) , \sin(\pi/2)] = [0, 1]$ later illustrated in \S\,\ref{section3.5}. The level of interpolation generally begins with $\delta_C$ which means with regard to the change required for compression $C$ to reach $C'$ otherwise conducting another $C$. We thereby constrict the domain of noise and fuzzy set within the product into binary sequence just like Fig.$\,$3, with respect to time $t$. This is by associating fuzzy set domain with pure maximum and minimum binary scale 00 and 11. One could identify this between layer exchange $\ell \leftrightarrow\ell'$, subsequently, $\ell' \leftrightarrow\ell''$ with respect to $t$. To show this from base to the continuous slope after pairwise combination of bits, we apply logical OR$\rightarrow$ AND/OR$\rightarrow$ AND, (inclined downwards left in Fig.\,3) to form a symmetry after an elapsed time of $t$-seconds. The integral is applied to the four layer construction of the FBAR algorithm. We of course, assume all conversion events including decompression occur in Hilbert space, since the use of quantum protocol at certain stages of the application is advantageous to the outcome of a high percent compression. The decompression layers involve the inverse function of psi values, or to be more specific, the wave carrying bits with length $\lambda$, further discussed in \S\,\ref{section3.5}. This is our standardized length measurement applicable to DB applications, resulting more efficient compressions compared to nowadays technology. To prove this, we illustrate and sufficiently explain a fuzzy qubinary register model throughout the following sections once the words ``\verb"Philip"'' and ``\verb"Baback"'' are compressed with a ratio of 3$:$1 in variable bit length limit, thereby decompressed successfully without losing data with our AND/OR method.
\vspace{-1mm}

\subsection{1st layer: String-to-binary conversion}
\label{sectio3.1}

We firstly bring about a two-cell attribute sample from a database as first and middle names (first names by an inheritable DB key) of the author:
\begin{center}
$\verb"Philip "\ \verb"Baback" $\ \ \ \ (\textit{source  string  database  input  example;  a  two-cell  example}) \\
\end{center}

\noindent Then we apply the exact conversion of the names' characters with space as a separator of the cells to binary values distributed as follows: \\

\noindent 01010000 01101000 01101001 01101100 01101001 01110000 \textbf{0010} 01000010 \\
\noindent 01100001 01100010 01100001 01100011 01101011\\

\noindent Sample code for conversion in a text field: \vspace{-1mm}
\scriptsize
\begin{flushleft}
\noindent---------------------------------------------------------------------------------------------------------------------------------\\
\texttt{0 \ Dim txt As String \\
1 \ Dim result As String \\
2 \ Dim strChar As String \\
3 \ Dim bin As String \\
4 \ Dim i As Integer \\
5 \ txt = txtAscii.Text \\
6 \ txt = Replace(txt, vbCr, "") \\
7 \ txt = Replace(txt, vbLf, "") \\
8 \ result = "" \\
9 \ 'For Loop statement is to convert and give its string-to-bin result\\
10 FOR i = 1 TO Len(txt) \\
11        strChar = Mid\$(txt, i, 1) \\
12        bin = LongToBinary(Asc(ch), False) \\
13        result = result \& Right\$(bin, 8) \\
14 NEXT i \\
15 txtBin.Text = result} \\
\noindent---------------------------------------------------------------------------------------------------------------------------------\\

\textbf{Note}: The \texttt{LongToBinary} function converts Long value into a binary string needed for preliminary steps of ASCII conversions before string-to-binary.
\end{flushleft}\normalsize

\noindent Obviously, the binary length recognition is of Rel.$\,$(7), such that the length of a word could be 8, 16, 32, ...-\emph{bit} type. In this layer, we consider the 8-\emph{bit} version for the laid out binary sequence due to its simplicity in being sorted by memory cells, thereby applying further binary and non-binary operations. The fixed word length conversion sometimes for unbalanced length of strings during conversiont applies, e.g., ``\verb"My Car"'' consists of 2 and 3 characters separated by a SPACE character. This length during its \emph{string-to-binary} and thereafter, \emph{binary-to-binary} (\emph{the encoding layers}) ASCII conversions must preserve the binary length limit just about approaching the final layers of data compression, e.g., 4th layer \emph{binary-to-string}, when \emph{measured}, must be clarified in quantity and experienced in \S\,\ref{section3.4.3}. In the following section, we experience the conversion of a 1-byte length after AND and OR application, return a $\frac{1}{2}$\emph{byte} = 4 \emph{bits} = 1 \emph{nibble}.

\subsection{2nd layer: A 2bit-AND parallel-to-OR compression technique}
\label{section3.2}

\noindent Upon this layer, from the initial string in the 1st layer, we simply map and stack in LIFO, a typical approach resulting a projective mapping of lower layer binary values to upper layer (2nd layer) values in form of stacked up nibbles. Each nibble represents the projected map per 1-\emph{paired bits} (2-\emph{bits} in total), neighboring each other from tail-to-head, and if from head-to-tail of the binary sequence (in case of not being LIFO), we later apply a \emph{string reverse function}, or in this case, \emph{binary reverse function} in our code. Nevertheless, the projection of 1st layer values comes to 3rd layer once the paired bits are geometrically focused, forming a sequence of focused values, i.e., a foci of paired bits.

\subsection{3rd layer: A pure pairwise Boolean detection technique}
\label{section3.3}

\normalsize The rationale incorporated into this stage merely advocates notations and their logic measured in quantum systems. The importance of including quantum noise indicators is due to making our algorithm capable to distinguish OR/AND from AND and OR separately during \emph{string-to-binary} conversions relative to their foci mappings per projection, from one binary plane to another, registered on a quantum computer. This layer, however, deals with projection of ANDed and ORed values in form of \emph{quantum bits topology} adjacent to \emph{classical bits topology } of information coming from the previous layers. It is consistent however, crucial in standardizing the very nature of how values are to be concatenated without a single loss of information between layers, as expected from~\S\ref{section1}. Ergo, from tail-to-head accumulation of bits delivered/projected to the 3rd layer for ``Philip Baback" emits
\begin{center}
 1000 1000 0000 0000 0000 0000 \textbf{0000} 0010 0000 0100 0000 0000 0000 \\
 1111 1011 1011 1011 1011 1001 \textbf{0010} 0011 1111 0111 1111 0111 0011 \\
 $k'$ \ \ \ \  $c'$ \ \ \ \ \ $a'$  \ \  \ \  $b'$  \ \   \  $a'$ \ \ \    $B'$ \ \ \ \ \ \  \ \ \ \    $p'$ \ \ \ \  $i'$ \ \ \ \ $l'$ \ \ \ \ $i'$ \ \ \ \ $h'$ \ \ \ \ $P'$ \\
\end{center}
\noindent Each nibble corresponds to ANDed output for a letter (nibble in the upper sequence) and ORed output of the same letter (a nibble in the lower sequence) but of reversed type. Therefore, a proper nibble expectation of the previous sequence becomes after reversing the reversed binary to its normal when projected onto the planes of the 4th layer. Two planes of a fixed size binary length are always ready to be concatenated for the 4th layer relative to a quantum noise inclusion. So, a pairwise boolean detection is necessary before any noise inclusion in the 3rd layer for further compression, because the total number of the current mapped bits in the 3rd layer are equal to the 1st layer number of bits, in this case, $12\times 8 = 96$ bits excluding SPACE. In fact, we say that we have just encoded the initial message, and compression is thus none or a $1$:$1$ ratio.

\noindent
\bigskip
\vspace{60mm}
\begin{flushleft} \hspace{-1mm}
\includegraphics[width=1.90mm, viewport= 0 0 10 10]{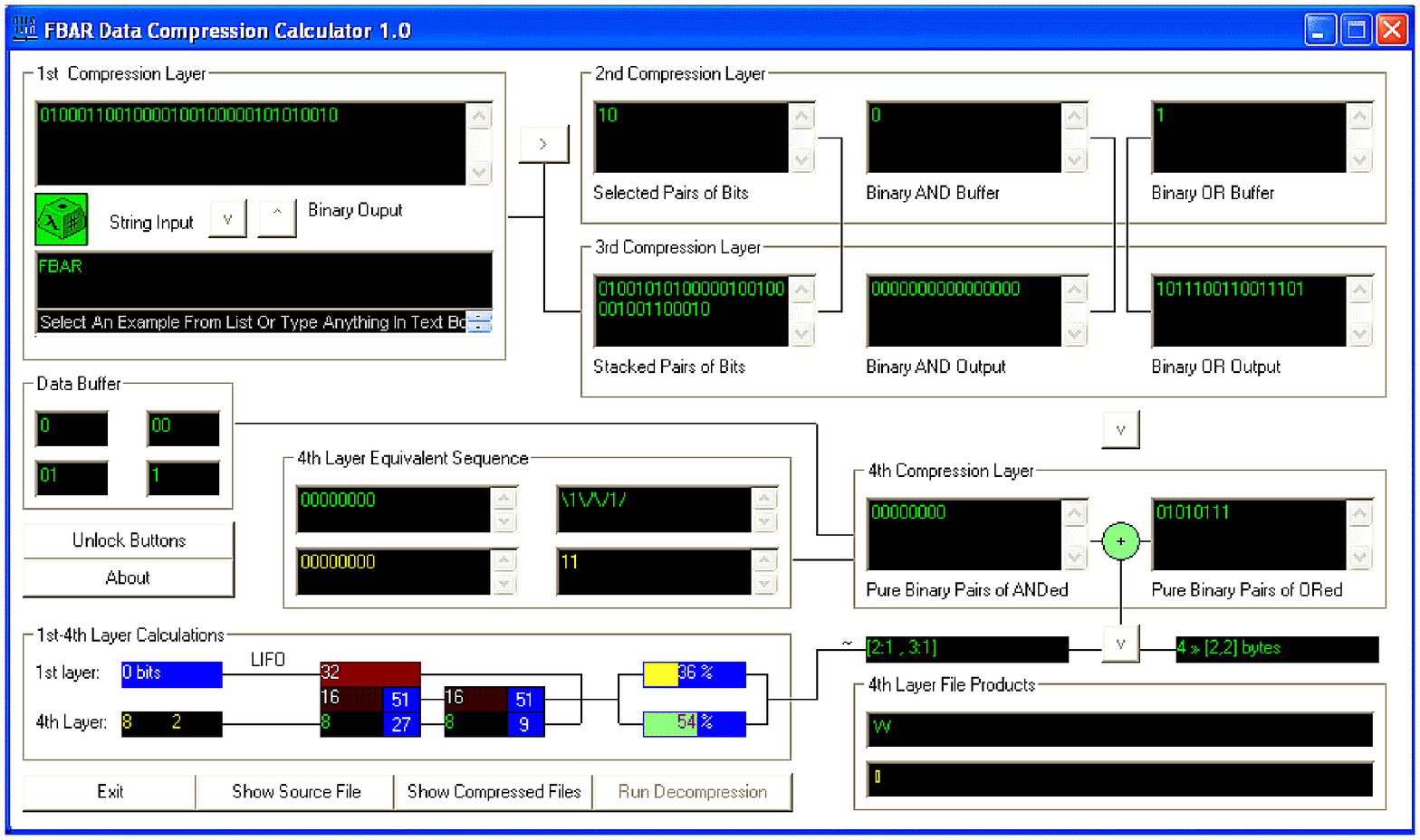}\\
\end{flushleft}

\noindent{\footnotesize{\textbf{Fig.$\,$4.} A screenshot of the FBAR Compression Calculator program, the trial version. The IV-layer compression is developed to serve data compression whilst data decompression to the four layers product is simulated for FQAR prior to FBAR classical binary conversions.

\bigskip
\normalsize

Therefore, we preemptively engage conformity of pure pairwise boolean detection with quantum probabilities quite illustratively deterministic in the set $\{01, 0, 1, 10\}$, prior to pattern recognition of the logic itself at the later stages of subsequent layers' initiations. This means that 00 converts to 0, and 11 to 1, and ``01 or 10" detected to be \emph{impure}, a direction to be of fuzzy type. The solution to this lies onto the way we allocate information based on polarity and bit position to our combinatorial logic which is quite fuzzy relative to being quite quantum in behavior, correspondingly. Once we encode the pure bits by simple coding, then we project our values in form of binary, fuzzy and quantum orders. This is subject to the 4th layer quantum and fuzzy binary inclusions.

\subsection{4th layer: A pure pairwise Boolean application using fuzzy quantum indicators }
\label{section3.4}

Using the example provided in \S\,\ref{section3.3}, and benefiting from symbolic string character substitutions, after reversing them via a binary reverse function when a final \emph{lifo sorting} is done, evaluating a byte length = 8-\emph{bits}, the \emph{quantum alphabet} for a LIFO binary sequence \emph{pure pairwise sorting} upon converted characters of the 3rd layer taken to the 4th layer appears as:
\begin{center}
\emph{ANDed pairs}: 00 00 00 $|$0$\searrow\rangle$ 00 $|\!\nearrow$0$\rangle$ \textbf{00} 00 00 00 00  $|$0$\nearrow\rangle$ $|$0$\nearrow\rangle$ \\
\emph{ORed pairs}: 10 $|$1$\searrow\rangle$ 11 $|$1$\searrow\rangle$ 11 10 $|\!\nearrow$\textbf{0}$\rangle$ $|\searrow\nearrow\rangle$ $|$1$\nearrow\rangle$ $|$1$\nearrow\rangle$ $|$1$\nearrow\rangle$ $|$1$\nearrow\rangle$ 11
\end{center}

\noindent We also included the \emph{ket psi} values, $|\psi\rangle$, whereby in this case, $bra$, or every logic state's dual (\emph{counterpart}) does exist once stored into some quantum memory. Ergo, the use of such substitutions becomes valid in an array of dual values.

Thus, the notion of \emph{endianness} or \emph{byte order} in detecting such duals becomes prioritized between two states of logic, no matter how precise in becoming fuzzy as a trigger to release a bit address that is needed for bit transactions between memory, registers and processor of the computer organization. The \emph{little-endian}/\emph{big-endian} priority over bit addresses in a fuzzy quantum register is readable when we store our individual indicators in form of dual values 01 and 10's. The most significant bit values are then neighbored to the most significant indicators from left to right dependant upon the address of where we have stored the duals and pure paired bits. Against the least significant ones, we reach an approximate $2^{\lambda(n)}$ trials to guess for the original data input, relative to the acquired length of compression i.e the AND-OR product. Hence, for the current product, a total length $\lambda(n) = \mathrm{ORed }+ \mathrm{ANDed} = 26 + 26 = 52$ bits excluding the raised 1-\emph{bit} flag operators could be evaluated. For a stacked set of compressed values in form of duals and paired bits $\lambda(n, \frac{1}{2}n) = \lambda(n) + \lambda(\frac{1}{2}n)=(17+4.5)+(22+2)= ]39, 45.5]$, and for beyond this compression, $\lambda(\frac{1}{n})= 1/45.5 \approx 0.024 \ bit$ appears out of a total 104 bits of source string input. The OR sign whereby concatenated with AND output before 2nd cache for this example served nibbles 0100 and 0000 for the string cell input, and when concatenated, form a SPACE character, raised by the a composite conversion of flag values from the previous layer into a $|\!\nearrow$\textbf{0}$\rangle\equiv\mathrm{Rev}(0010)=0100$ and \textbf{00}$\, \equiv\mathrm{Rev}(0000)=0000$ respectively.

In this layer, however, the AND/OR sorting gave us a total of 20 bits. In this process, the first NULL is skipped or ignored by the algorithm, and for the second, the flag is already raised, or, an amount of $\approx 0.58$ input data is compressed per binary flag exclusion.

In this pure pairwise binary sequence integration, the FBAR program (Fig.$\,$4), thereby accesses memory location (address) to categorically via LIFO (\emph{last-in-first-out}) with conversive functions including For Loops convert and compress data in a classical process mimicking a fuzzy qubinary computation.

The process firstly launches a further possible compression on FBAR through quantum indicators. For example, the equivalent of 01, could be encoded as ``/'' as a string length = 1, or, 1-\emph{byte character}= 8 \emph{bits} in the FBAR program due to the program compiler limitations. Secondly, the process stores data as file binary signals (this is not related to the notion of a \emph{binary file} in structure) against probabilistic combined signals of volumetric binary space within the compressed textual files. In other words, according to Fig.\,5, by representing the stages or layers of FBAR data compression algorithm, the application evolves in form of storing indicators ``/" and ``\textbackslash'' as ``1''and ``0'' closures respectively. These indicators are to be inserted or substituted i.e., \emph{strongly programmed as code-dependant} by a detected \emph{character-by-character} search, in form of zero-byte files under the ``01'' and file directories.
\bigskip
\vspace{132mm}
\begin{flushleft} \hspace{-1mm}
\includegraphics[width=1.65mm, viewport= 0 0 10 10]{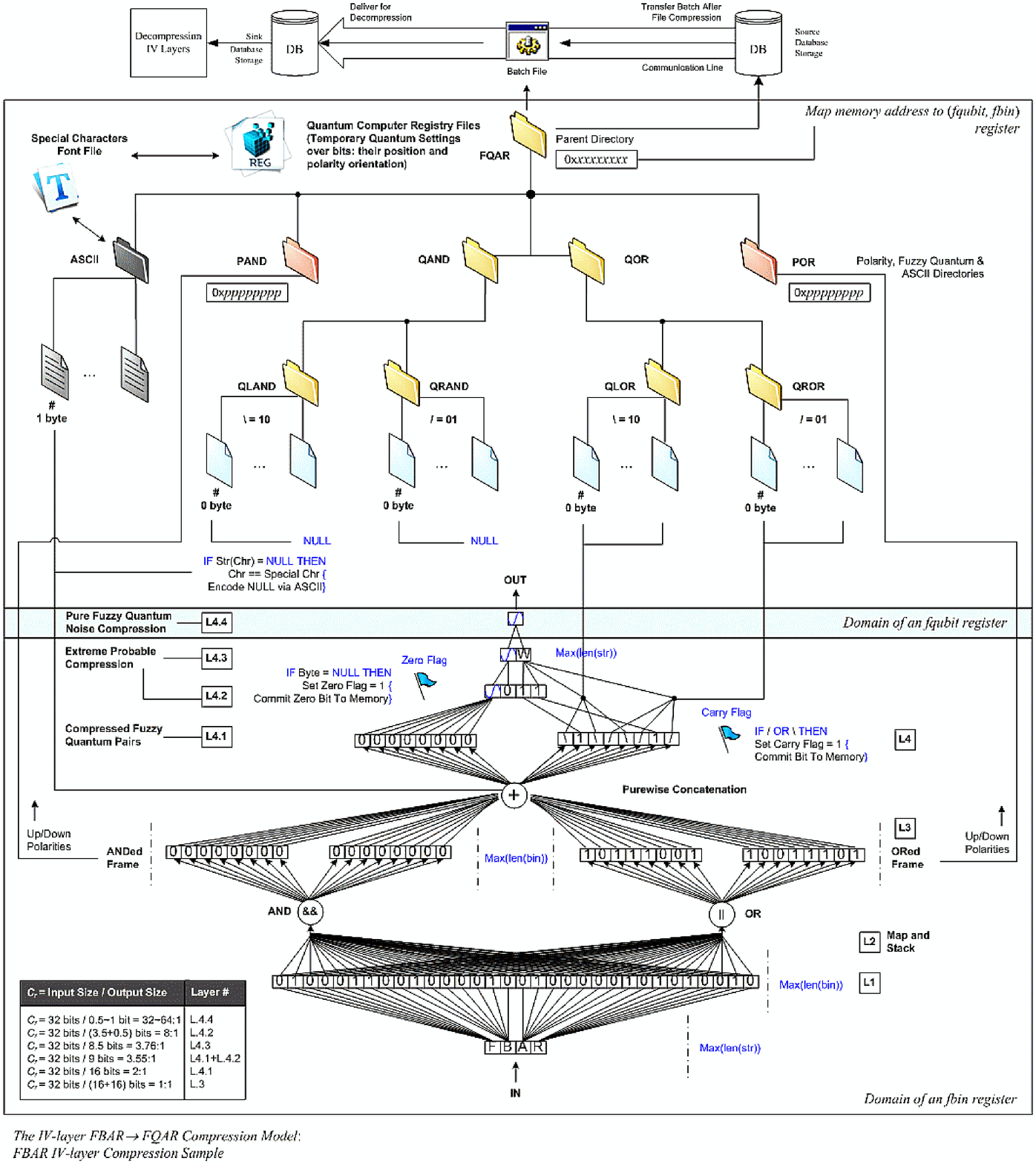}\\
\end{flushleft}

\noindent{\footnotesize{\textbf{Fig.$\,$5.} To have an illustratable pattern on data compression, we tried ``\texttt{FBAR}" in form of a text input. The result from one layer to another, mathematically confirms the deportment of \emph{string-to-binary} mapping, and thereby \emph{binary-to-binary}, (\emph{the encoding layers}) ASCII conversions preserving the binary length limit just about approaching the final layers of data compression, e.g., 4th layer \emph{binary-to-string}, with storing history of ASCII code conversions in form of zero-byte files, their properties and virtually configurable attributes containing a particular level of layer conversion result. This is essential for decompression to access historical data on the compressed version with a minimum size of occupiable HDD space.  \\

\long\def\symbolfootnote[#1]#2{\begingroup%
\def\thefootnote{\fnsymbol{footnote}}\footnote[#1]{#2}\endgroup}

\normalsize
The comprehensive geometric and mathematical illustration on Fig.$\,$4 trying a string, ``\texttt{FBAR}'', is given in Fig.$\,$5. The concatenated AND-OR string sequence from the 4th compression layer is separated from one another during the decompression process explained below, symmetrically.

The vertices between layers indicate \emph{data types} as binary, expected from the last layer which is a combination of string and fuzzy qubinary character data types. The foci in form of multiple operators are quite discrete in essence to the binary conversions from \emph{string-to-binary} and vice-versa conditions. This is mathematically abstracted via \emph{convert function} in our program code.

The foci could be classed according to Fig.$\,$5, as AND, OR and Concatenation operators denoted by $\texttt{\&\&}$, $\texttt{||}$ and $\texttt{+}$ symbols respectively, incorporated between layers I-IV of the data compression phase.

In the ``\texttt{FBAR}" compression example provided, 4 string vertices are mapped to 32 binary vertices forming the 1st-to-2nd layer conversion, since we know 1 character value is equivalent to 8 bits, hence 4 characters generate 32 bits based on ASCII binary and character conversions. Subsequently, 32 binary vertices (or sequence) are focused in operation into two vertices and thereafter 16 parallel to 16 vertices as a subfunction to the former, or
\begin{equation*}
\operatorname{Convert}(f(v_{\texttt{str}}))\mapsto f(v_{\texttt{bin}})= \wedge\vee   \longrightarrow \ell (C, t) \ \ \ \ \ \ \ \ \ \ \ \ \ \ \ \ \ \ \ \ \ \ \ \ \ \ \ \ \ \ \ \ \ \ \ \ \ \ \ \ \ \ \ \ \ \ \ \ \ \ \ \ \ \ \ \ \ \
\end{equation*}
\vspace{-4mm}
\begin{equation*}
\mathrm{when} \ f(\ell) \mapsto f (\ell') \ \mathrm{then} \  \operatorname{Convert}(f(v_{\texttt{str}}))\mapsto  v_{\texttt{bin}}) =  f(\operatorname{Convert}(g(\wedge\vee  v_{\texttt{bin}})) = \ \ \ \ \ \ \ \ \ \ \ \ \ \ \ \ \ \ \ \ \ \ \ \ \ \ \ \ \ \ \ \ \ \ \ \ \ \ \ \ \ \ \ \ \ \ \ \ \ \ \ \ \ \
\end{equation*}
\vspace{-4mm}
\begin{equation}\label{9}
\{\frac{1}{2}bin +  \frac{1}{2}bin'\} \vee \{\frac{1}{n} bin + \frac{1}{2} bin'\} \vee \{\frac{1}{n} bit + \frac{1}{m} bit'\}= \ell (C, t'+1) \ \ \ \ \ \ \ \ \ \ \ \ \ \ \ \ \ \ \ \ \ \ \ \ \
\end{equation}\vspace{-1mm}

\noindent The previous relation strictly confines itself within the norms of \emph{binary rankings} for every paired bits leading to their qubit definition discussed in \S\S\,\ref{section2} and \ref{section3} (apart from their subsections) covering the fuzzy-binary concept made on Figs.$\,$2, 3 and 5. In this relation, the $\wedge\vee $ denote AND, OR operators to AND and OR values from one binary plane to another. This encodes the binary sequence into two columns illustrated in Fig.\,5. The transformations for every co-occurring encode between AND and OR columns, results in a $(C, t\rightarrow t'+1)$, which means that the eventual outcome of two parallel conversions (specified by $\operatorname{Convert}$) on a set of string vertices $v_{\texttt{str}}$ (the use of function $f$ is to identify its sum or set properties, e.g., its cardinality and data type), from layer $\ell$ to $\ell'$ is a compressed binary sequence for OR as ORed values located as vertices of binary type or $v_{\texttt{bin}}$, and its parallel column which is ANDed or, $\{\frac{1}{2}bin +  \frac{1}{2}bin'\}$.

An uncorrelated result of ANDed and ORed sequences, result in different lengths given by $n$, $m$ \emph{singularities}\symbolfootnote[1]{Here, as singular points near absolute 0, once reached, the carrying signal degenerates; see, also Ref.~\cite{45-Knopp}.} commencing with the \emph{trivia risings} of 1/2 data between two sequences of parallel binaries ending up with singular forms of quantized bits with possible fuzzy combinations. The extreme uncorrelatedness, however, applies to fragmented bits carrying bit values known as dots (next subsection). Such bits are quite fragmented when committed to some quantized signal, a hypothesis yet to be tested for data decompression purposes. All three probable outputs are treated as values of maximum compression $C$, whereby its elements abide by the limits of sequential computation, hence the use of $\vee$ between outputs is of conventional use to correlate them when decoded for decompression $C'$. The two vertices related to basic OR and AND conversion operations are in fact a foci of AND and OR switches, and thus are applied simultaneously at time $t=t'=0$, where $t$ is parallel to $t'$ during binary-to-binary computation. Obviously, the remaining foci are mapped from the AND/OR mapping layer to the 4th layer in terms of a single vertex, which represent a focus to all subsequent binary vertices as the subfunction of the latter layer (layer III). This time, since binary pairs are detected for pure states neighboring those impure or \emph{fuzzy quantum pairs}, ergo comes in the notion of the 4th layer quantum and non-quantum or binary, 1/2 state binary (1/2 bit) quantum integration, generating a combinatorial string of ASCII characters and fuzzy quantum units. Such data integrity between, e.g., two databases in terms of fuzzy quantum states are explained in the following subsections whereas the decompression phase is debated conversely in \S\,\ref{section4}.

\subsection{4th layer: A quantum noise inclusion with fuzzy logic indicators}
\label{section3.5}

\normalsize
We symbolically confined our storage system for quantum noise indicators in terms of ``/ and $\backslash$" denoting states of `$01$ and $10$' adjacent to their binary sequence from an ORed and ANDed layer. The compressed content of this data at the upper edge of the sub-layers of the 4th layer is fragmented between upper half and lower half of a quantized signal. We suppose that our system possesses qubit registers for a quantum computation. 

These states of compression could be expressed as, e.g., an ordered form of a \emph{fragmented fractal compression sequence}, or simply, a \emph{compressions' sequence} $C_1 > C_2 > C_3 > \ldots > C_n$, occupied by a set of data objects (\emph{binary dots}) in a quantized signal structure, thereby generating $C_1\gg C_n$. One could conceive these dots as inner product of a \emph{bra-ket psi} value within the Hilbert space which does precisely venture a dual space $\mathcal{H}^\textbf{*}$ to its corresponding dots as well. The signal is posed (or superposed) by at least two incoming-propagating electromagnetic waves across quantum circuit denoting a definite quantum topology.

In Fig.$\,$6, the classical boolean values of our algorithm is projected in form of differentiable dots from a 2D-plane where classical bits live in a \emph{topologically-concentrative loci} of the same dots pattern mapped to a 1D dot occupant (just a dot) for a spatially quantized signal locus to a infinite-dimensional norm or Hilbert space, i.e. independent of time in case of two quantized signals revolting with synchronicity (superposed) or $t_i=t'_i$.

The dot value (a \emph{sub-sub-signal}) denotes certain properties of a compressed bit in terms of its: 1- \emph{logic state}, 2- \emph{position}, 3- \emph{address}, and 4- \emph{polarity}. The properties' list could be fragmented into at least 2-\emph{fqubit registers} clocked with synchronicity: one representing the first two properties of dots and the other, for the remaining properties given to its parallel signal. Asynchronous cases with wave latency could be idealized between planes of projection physically illustrated for the decompression phase of the algorithm, subject to the following paragraphs in virtue of Fig.$\,$7. The late \emph{wavelet} arrival time is denoted by $\tau$ which is the delayed time encountered, measured by its delayed wavelength or $\lambda'$ which is equal to $\lambda$ fragments divided by the \emph{speed of light} or $\tau = \lambda /2n c$, where $c = 299,792,458 \ \mathrm{m s^{-1}}$. The more $n$ carried, the least delayed wavelet which means $n$ bits of information or dots are ahead of a probable latency event occurrence, and now ready for projection.

The current method must at least corroborate with the findings of Zhang \emph{et al.}, Arimondo \emph{et al.} and Phillips~\cite{{36-Zhang et al.,{38-Arimondo et al.},{39-B. Phillips}}}, dealing with `robust probabilistic quantum information processing' with \emph{atoms}, \emph{photons} and \emph{atomic ensembles }subject to future reports relative to FBAR as FQAR. The extreme compression of data is indeed exponentially efficient once accomplished for a physically condensable atomic frames representing a qubit register accumulating differentiable planes as 2D-stacks of dual and singular bit positions in the ANDed and ORed binary sequences. This is illustrated in the upper quantized signal of Fig.$\,$6. The current concept eventually concludes with the proportionality
\vspace{1mm}
\[\frac{projected \ compressed \ data}{time \ needed \ for \ projection} \propto atomic \ condensation \ frequency \ ,  \]
\[ \because P=P^2 \ , \ \therefore \frac{P(C(classical \ bit \ property))}{P^2(C) \times t} \longrightarrow C(C) = \]
\[ C(fqubit \ property) \longrightarrow \frac{C(C_1, C_2, C_3\ldots,C_n)}{t}, \]
\vspace{-3mm}
\noindent deducing
\begin{equation} \label{10}
\therefore n\frac{\lambda}{2}\stackrel {P^2\rightarrow P}{\longrightarrow}\frac{\lambda}{2}=  n\frac{\lambda}{2} \backsim \frac{\lambda}{2} \equiv n:1
\end{equation}

\noindent which is defined in a quantitative manner for a wavelength $n\frac{\lambda}{2}\backsim \frac{\lambda}{2}$ denoting a standing projected wave propagation between two lattice atoms of some 1D-qubit register. This must be conditioned for lossless applications no matter the degree of compression, the quality of data must be preserved for every $C_i$ composition (physical representation) in practice. The projective mapping gives a ratio $n$:$1$ per signal frequency, expectably. The given proportionality becomes explicitly harmonious with its projector of compressed data with a projection $P$ between non-quantized and quantized planes harboring incoming signals that obey $P^2$ for each co-occurring $P$ between planes, making $P^2$ = $P$ at time $t=t+1-1$ or $t=t'$ relativistically.\symbolfootnote[1]{Noteworthy to emphasize on the concept of time translation to be of Lorentz transformation, more of a \emph{symmetric configuration} between the clocked signals.} This forms a multidimensional compression or, $C(C_1, C_2, C_3\ldots,C_n)$ for the von Neumann's \emph{mean time} given, and thus measured in Hz or $\mathrm{s}^{-1}$, until the signal leaves the \emph{n}-fqubit register system. This measurement is merely applicable to the FQAR system whilst data gets compressed from one lower norm to its upper. The final product of all upper norms of the 4th layer, is then measured in terms of \emph{fqubit per second} or \emph{fqubit-rate} expectably. The rate reflects values of attuned oscillated signals when in a projective position. This is merely evaluated once we attain a level of a \emph{path-dependant decompression}, whereby the ``condensation frequency" is measured for asynchronous cases forming a \emph{hysteresis} Load/Unload data condition illustrated in Fig.$\,$7 \emph{a}); assuming that all noise transmissions are synchronized per co-occurred projections (at least $2P$'s).

\smallskip

\vspace{94mm}
\begin{flushleft} \hspace{-1mm}
\includegraphics[width=1.4mm, viewport= 0 0 10 10]{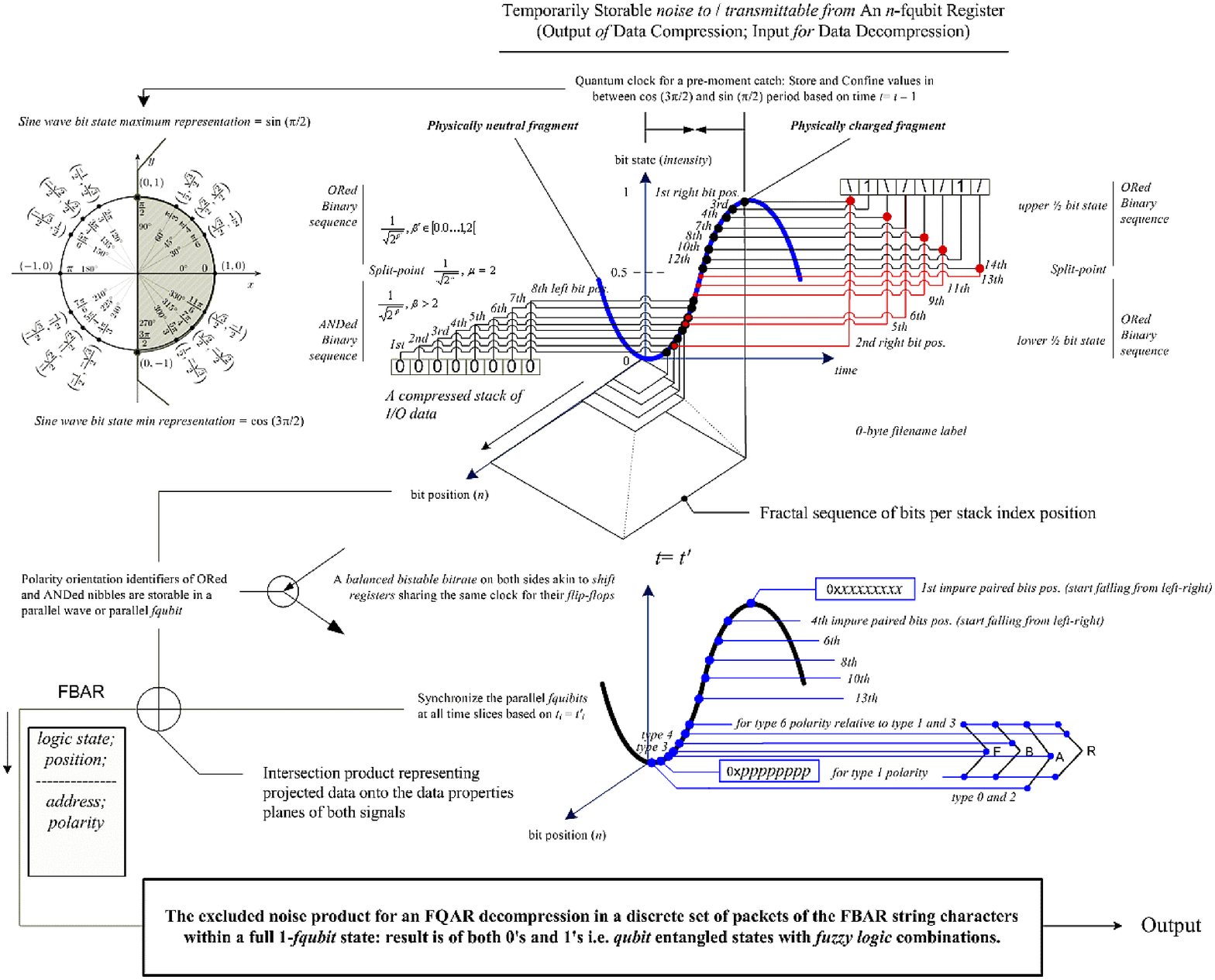}\\
\end{flushleft}
\vspace{-2mm}
\noindent{\footnotesize{\textbf{Fig.$\,$6.} Two fqubit registers taking physically-intersected datasets as temporarily storable compressed data for maximum compression at the upper 4th layer of the FBAR algorithm. }

\long\def\symbolfootnote[#1]#2{\begingroup%
\def\thefootnote{\fnsymbol{footnote}}\footnote[#1]{#2}\endgroup}
\normalsize
\subsubsection{The 4th layer seclusive proposal}
\label{section3.4.1}

\noindent \emph{A conditionally seclusive\symbolfootnote[1]{The idea is to practically implement our grounds of proposal in a secluded experimental environment, quite secured for precise measurement, thus reaching a reliable verdict for ensuing statistics from computational physics experiments proving FBAR as FQAR.} proposal:}----------- In Fig.$\,$7\emph{a}) we show two incoming fqubits carrying dots with their properties. The challenge is to address the registered dots per 1D-qubit whilst fuzzy indicators for a cyclic site of lattice reserved. We register such dots by pumping from an upper 1D-lattice to a lower one, quite descriptive in Arimondo \emph{et al.} notes~\cite{38-Arimondo et al.} and illustratively depicted in the top lattice site of Fig.$\,$7\emph{a}). The lattice consists of trapped atoms, a sample prepared by ``optical pumping" which by a series of absorption-emission cycles prepares all atoms to the same quantum state, for example, the state corresponding to the logical 0, pp.$\,$5-6~\cite{38-Arimondo et al.}. This is recognizable in Fig.$\,$7\emph{a}) for the projected waves. A well-defined approach via the atomic Bose-Einstein condensate is applicable to convert 1D-lattice to stable 2D and 3D arrays huge atomic quantum register of up to 100,000 qubits, with a single atom at each lattice site.
\vspace{156mm}

\begin{flushleft} \hspace{-1mm}
\includegraphics[width=1.76mm, viewport= 0 0 10 10]{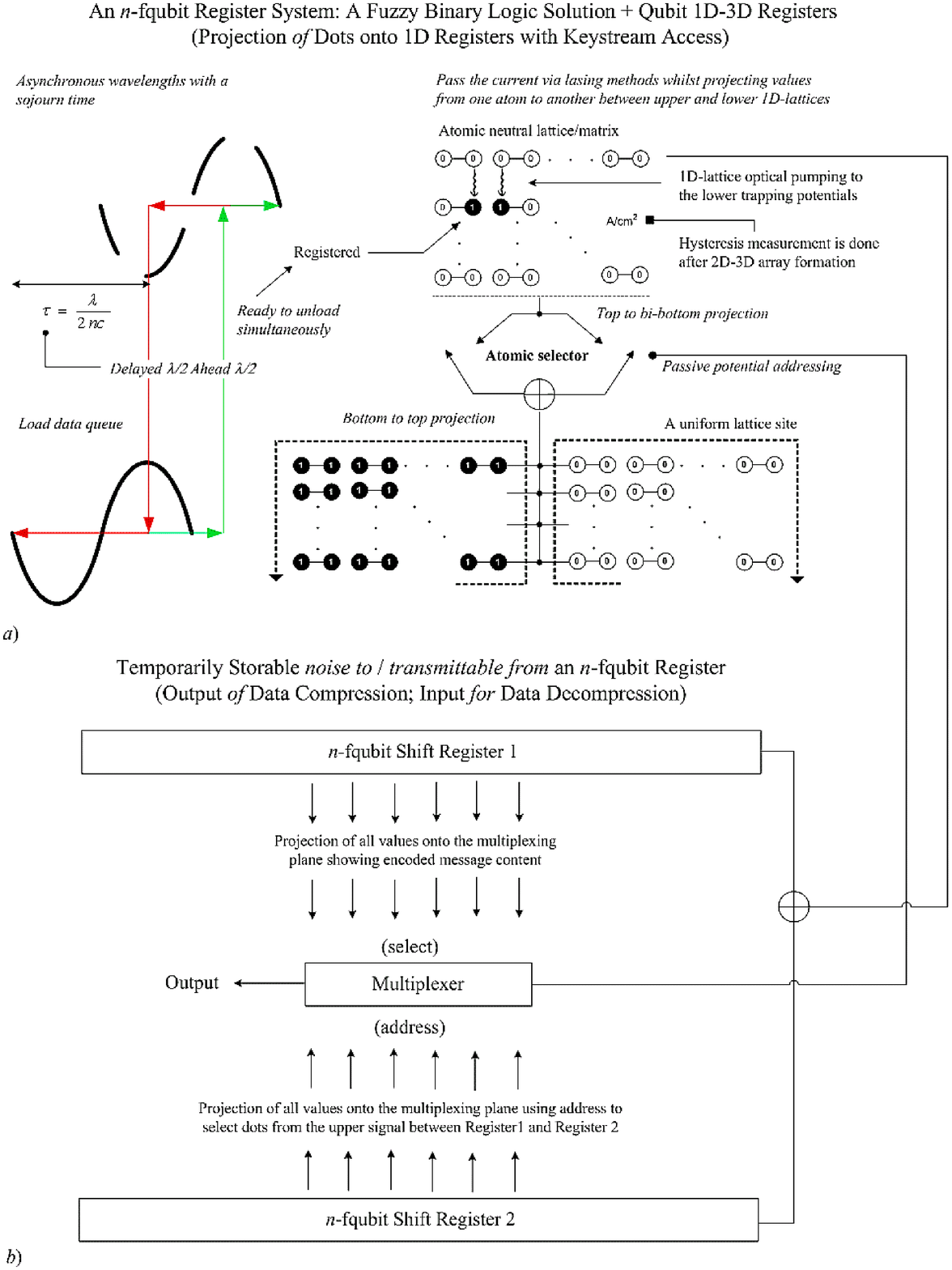}\\
\end{flushleft}
\vspace{-2mm}
\noindent{\footnotesize{\textbf{Fig.$\,$7.} (\emph{a}) An \emph{n}-fqubit register system with latency overcoming options between clocked quantized signals. Upper and lower band limits of each signal act as carriers of compressed information refreshed by two lower middle atomic lattices relative to the top lattice interfaced by an \emph{atomic selector}, which preserves the contents of projected signals per se. This is of lossless method promoting FBAR to FQAR; (\emph{b}) A 2\emph{n}-fqubit multiplexer generator taking in signals from the upper image for data decompression purposes. }
\bigskip

\normalsize
Such atoms are locus in representing \emph{ket psi} $|\psi\rangle $= 0 and 1 values for a micro-potential of $\lambda/2$=0.5 $\mathrm{\mu m}$ in the 1D-to-1D lattice: corresponding to its row vector \emph{bra} values are established when optical pumping occurs for \emph{transposing kets} i.e. 1D-to-1D dot projection. An example of using ultracold atoms in a optical lattice is discussed by Arimondo \emph{et al.}, where the challenge is in selecting a single atom for addressing purposes just the one in use by our model, the \emph{atomic selector}. Dieter Meschede and his group in Bonn have shown that by using magnetic field gradients it is possible to select single atoms separated by a few lattice sites and it remains to be seen whether such an advanced level of control can be extended to the quantum register of a Mott insulator~\cite{41-Schrader et al.}. Other possibilities to address single atoms are currently pursued by the group of William D. Phillips at NIST (USA). We propose the fqubit multiplexer equivalent to an atomic selector for the purpose of selecting the right logical values of a looping field between two registered dots: the \emph{bi-bottom lattice sites} in Fig.$\,$7\emph{a}) aiming at its objectives, simplified into Fig.$\,$7\emph{b}) as an \emph{fqubit multiplexer generator}. These dots represent all data in the upper and lower waves (fqubits) between compression and decompression phases of FBAR and FQAR. We suggest the selection of logical values between sites to be projective from bottom-to-top lattice sites and vice versa, such that every scan corresponds to 1-to-$n$ selections similar to \emph{passive matrix addressing} as an addressing scheme used in earlier LCD displays; in our case $m + n$ control signals are required to address an $m \times n$ lattice site (display). An atom, here for a dot value which could be any compressed data index contained by it (Fig.$\,$6), in a passive matrix, must maintain its state without active driving circuitry until it can be refreshed again, i.e. the ``bi-bottom lattice sites" refreshed periodically to normalize against any data loss contiguously. The selected micro-potential at selected row $i$ and column $j$ for the preserved data queue upon a dot in the lattice could be given in terms of
 \[ \sum V_{\mu_{ij}} = \sum V_{\mu_{\mathrm{sel} | \mathrm{unsel}}} - \sum V_{\mu_{\mathrm{on} | \mathrm{off}}}  \]
\noindent where this relation corresponds to both \emph{unselected} and \emph{selected} micro-potentials $V_{\mu_{\mathrm{sel}}}$ and $V_{\mu_{\mathrm{unsel}}}$, as if atoms with the same $n{\lambda/2}$ distances (separations) are addressed for dots of the top lattice site correspondingly. An on-brighten atom from the ``bi-bottom lattice sites" holding a dot value for the top lattice site, corresponds to $V_{\mu_{\mathrm{on}}}$  and an off-switched  corresponds to $V_{\mu_{\mathrm{off}}}$  potential. The use of sum $\sum$ is symbolic, denotes that any matrix operation could apply for $V$'s and their indices $i$ and $j$, as far as a micro-potential selection including its negation are satisfied. Once the selection and the refreshing loop solution (uninterrupted data supply) are in position, \emph{keystream} for data addressing, polarity and position would be less complex to select and thereby decompress the initial message. The selection would be from lower wave for the upper one to decipher (4th maximum layer to 1st), where each contain specific dots corresponding to a compressed value per 0.5 sate occurrence. We thus promote the concept of FBAR to FQAR for the mere purpose of an extreme compression plausible on a physical scale with high grade of data delivery i.e. application level. -----------

\subsubsection{The 4th layer bit asymptotes with a mean value of half-bit representing fuzzy quantum noise state}
\label{section3.4.2}

Let a fragmentary ``bit partitioning'' or ``an equipartition fragment order of a quantum signal" denote an asymptote of a Boolean bit value in terms of:

\begin{equation}\label{11}
\frac{1}{n} bit \rightarrow \min(bit) = \frac{1}{m} bit' \ , \ \frac{1}{n} bit \rightarrow \max(bit) = \frac{1}{2} bit'
\end{equation}

\noindent In addition, let the formulated asymptote satisfy conditions of detected values of fuzzy logic indicators in virtue of pattern polarity, exemplified in \S\,\ref{section3.3}. Thus
\begin{equation}\label{12}
\frac{1}{n} f(\nearrow\searrow) \rightarrow \max (bit)= \frac{1}{4} \{ \mathop {\nearrow}\limits^{\underline{1{\rm fqbit}}}, \mathop {\searrow}\limits^{\underline{1{\rm fqbit}}} \} \equiv \frac{1}{8} \, 2 \, paired \, bits \
\end{equation}

\noindent Therefore, the arrow indicators illustrated in \S\,\ref{section3.3}, become abstracted before the AND/OR application. The paired binaries are encoded in terms of \\
\scriptsize
\begin{flushleft}
\noindent---------------------------------------------------------------------------------------------------------------------------------\\
\texttt{0 \ START FBAR FUNCTIONS \\
1 \ FUNCTION AND\_PureOutput() \\
2 \ Dim c1, c2, ANDBuff(1) As String \\
3 \ Dim i, j, m, n As Integer \\
4 \ FOR m = 1 TO Len(binaryANDed\_Stack) - 1 \\
5 \ FOR n = 1 TO Len(binaryANDed\_Stack) \\
6 \ c2 = Mid(binaryANDed\_Stack, m, 1)  ' This is the (nth character - 1) of string \\
7 \ NEXT n \\
8 \ ANDBuff(0) = c2 \& c1  ' Display in form of binary pairs before AND/OR \\
9 \ NEXT m \\
10 SELECT CASE buff(0) \\
11 CASE "00"   ' This is a pure Low-level logic indication \\
12 ANDBuff(1) = "0" ' This buffer is of string type \\
13 CASE "01"   ' This is an impure state; a fuzzy-quantum inclusion of L-to-H level logic \\
14 ANDBuff(1) = "0" '01--> quantum indicator $\slash $ a prime(0) $==$ closure 1 from base 0 in [01] \\
15 CASE "10"   ' This is an impure state; a fuzzy-quantum inclusion of H-to-L level logic \\
16 ANDBuff(1) = "1" '10--> quantum indicator $ \backslash $ a prime(1) $==$ closure 0 from base 1 in [10] \\
17 CASE "11"   ' This is a pure High-level logic \\
18 ANDBuff(1) = "1" \\
19 END SELECT\\
20 PureBinaryPairs\_AND.Text = buff(1) \& PureBinaryPairs\_AND.Text \\
21 END FUNCTION}  \\
\noindent---------------------------------------------------------------------------------------------------------------------------------\\
\texttt{22 FUNCTION OR\_PureOutput() \\
23 \ Dim c1, c2, ORBuff(1) As String \\
24 ... 'we repeat lines \# 2 to 8 except with binaryORed\_Stack  \\
25 ORBuff(0) = c1 \& c2 ' Display in form of binary pairs before AND/OR application \\
26 NEXT m \\
27 SELECT CASE buff(2) \\
28 CASE "00"   ' This is a pure Low-level logic indication \\
29 ORBuff(1) = "0" \\
30 CASE "01"   ' This is an impure state; a fuzzy-quantum inclusion of L-to-H level logic \\
31 ORBuff(1)= "1" '01--> $ \slash $ a prime(1) $==$ closure 0 from base 1 in [01] \\
32 CASE "10"   ' This is an impure state; a fuzzy-quantum inclusion of H-to-L level logic \\
33 ORBuff(1) = "0" '10--> $ \backslash $ a prime(0) $==$ closure 1 from base 0 in [10] \\
34 CASE "11"   ' This is a pure High-level logic \\
35 ORBuff(1) = "1" \\
36 PureBinaryPairs\_OR.Text = ORBuff(1) \& PureBinaryPairs\_OR.Text \\
37 END SELECT \\
38 END FUNCTION} \\
\noindent---------------------------------------------------------------------------------------------------------------------------------\\
\texttt{39 FUNCTION AND\_OR\_Output() \\
40 Dim c1, c2  As String \\
41 Dim i, j, m, n As Integer \\
42 FOR m = 1 TO Len(binary.Text) - 1 \\
43 FOR n = 1 TO Len(binary.Text) \\
44 c1 = Mid(binary.Text, n, 1)  ' This is the nth character of the string \\
45 c2 = Mid(binary.Text, m, 1)  ' This is the (nth character - 1) of the string \\
46 NEXT n \\
47 binaryPairs.Text = c1 \& c2 ' Display in form of binary pairs before AND/OR \\
48 binaryAND.Text = c1 AND c2 \\
49 binaryOR.Text = c1 OR c2 \\
50 NEXT m \\
51 END FUNCTION \\
52 END FBAR FUNCTIONS} \\
\noindent---------------------------------------------------------------------------------------------------------------------------------\\
\end{flushleft}
\normalsize
\noindent We have demonstrated that the encoding for abstracted versions of binary into fuzzy quantum logic is possible, e.g., a 2-bit combination halved becomes 1, and 1 bit halved becomes 0.5, and ... So, instead of a textual output, one could specify the equivalent form in fuzzy binary logic. However, there are certain limits to be concerned of during conversions addressing word length and ASCII code variables.

\subsubsection{The 4th layer removal of fixed word length limitation and ASCII conversions}
\label{section3.4.3}

We conclude the state of the lossless data compression by removing the fixed word length for unbalanced length results on a fuzzy quantum binary level. This is how it goes in terms of a VB semi-pseudocode:

\scriptsize
\begin{flushleft}
\noindent---------------------------------------------------------------------------------------------------------------------------------\\
0 \texttt{\ START FBAR COMPRESSION \\
1 \ Execute \emph{string-to-binary}... \\
2 \ OUTPUT: Binary Product. \\
3 \ \emph{binary-to-binary}... \\
4 \ Encoded Once \\
5 \ Encoded Twice \\
6 \ Encoded... \\
7 \ OUTPUT: Binary Product. \\
8 \ 'The following reflects a section of layer \# 4 of the FBAR algorithm. \\
9 \ Measure \emph{binary}... \\
10 Dim i(0 To 1), j As Integer \\
11 \emph{len}(\emph{word}) = \{8, 16, 32, 64\} \\
12 \emph{max}(\emph{len}(\emph{word})) = \emph{len}(\emph{word}) $\ast$ 2 \\
13 SELECT CASE \emph{len}(\emph{bin}) \\
14 CASE 1 TO 7, 9 TO 15, 17 TO 31 IS $<=$ \emph{len}(\emph{word}) OR CASE IS $>=$ \emph{len}(\emph{word}) $\ast$  2 \\
15 IF \emph{max}(\emph{len}(\emph{bin})) $<$ \emph{max}(\emph{len}(\emph{word})) THEN \\
16 \emph{max}(\emph{len}(\emph{bin})) = \emph{max}(\emph{len}(\emph{bin})) - \emph{len}(\emph{word}) \\
17 i(0) =  \emph{max}(\emph{len}(\emph{bin}))  '0 is an array index instantiation; store new value to i  \\
18 \emph{max}(\emph{len}(\emph{bin})) = i(0) + \emph{len}(\emph{word})  'return and assign new length value \\
19 ELSE \\
20 i(1) =  \emph{max}(\emph{len}(\emph{bin})) '1 is an array index instantiation \\
21 \emph{max}(\emph{len}(\emph{bin})) = \emph{max}(\emph{len}(\emph{bin})) - \emph{max}(\emph{len}(\emph{word}))\\
22 END IF  \\
23 FOR j = 1 TO \emph{max}(\emph{len}(\emph{bin})) 'this is a For Loop \\
24 \emph{bin} = \emph{bin} + "01..."  'Catenate with some 0's and 1's up to max new bin-length  \\
25 NEXT j \\
26 END SELECT \\
27 \emph{binary-to-string} \\
28 String Product... \\
29 OUTPUT: Store Product. \\
30 HALT FBAR COMPRESSION} \\
\noindent---------------------------------------------------------------------------------------------------------------------------------\\
\end{flushleft}
\normalsize
Line \# 14 of the semi-pseudocode interpreted as the current length of binary sequence, if $<$ 8 and $>$ 0, or, $>$ 8 and $<$ 16, or, $>$ 16 and $<$ 32 and ..., which must be equal to the same length subtracted by the relevant fixed word length (being of e.g., word length 8, if and only if the condition for the binary sequence length is $>$ 8 and $<$ 16), such that its new result is thereby added to the previous result on the total binary length \emph{max}(\emph{len}(\emph{bin})). Therefore, totalling a devisable binary sequence with no remainder when converted to proper string characters, quite compatible with fixed word length string conversion satisfying a proper data compression.
\smallskip
\vspace{63mm}
\begin{flushleft} \hspace{-1mm}
\includegraphics[width=1.39mm, viewport= 0 0 10 10]{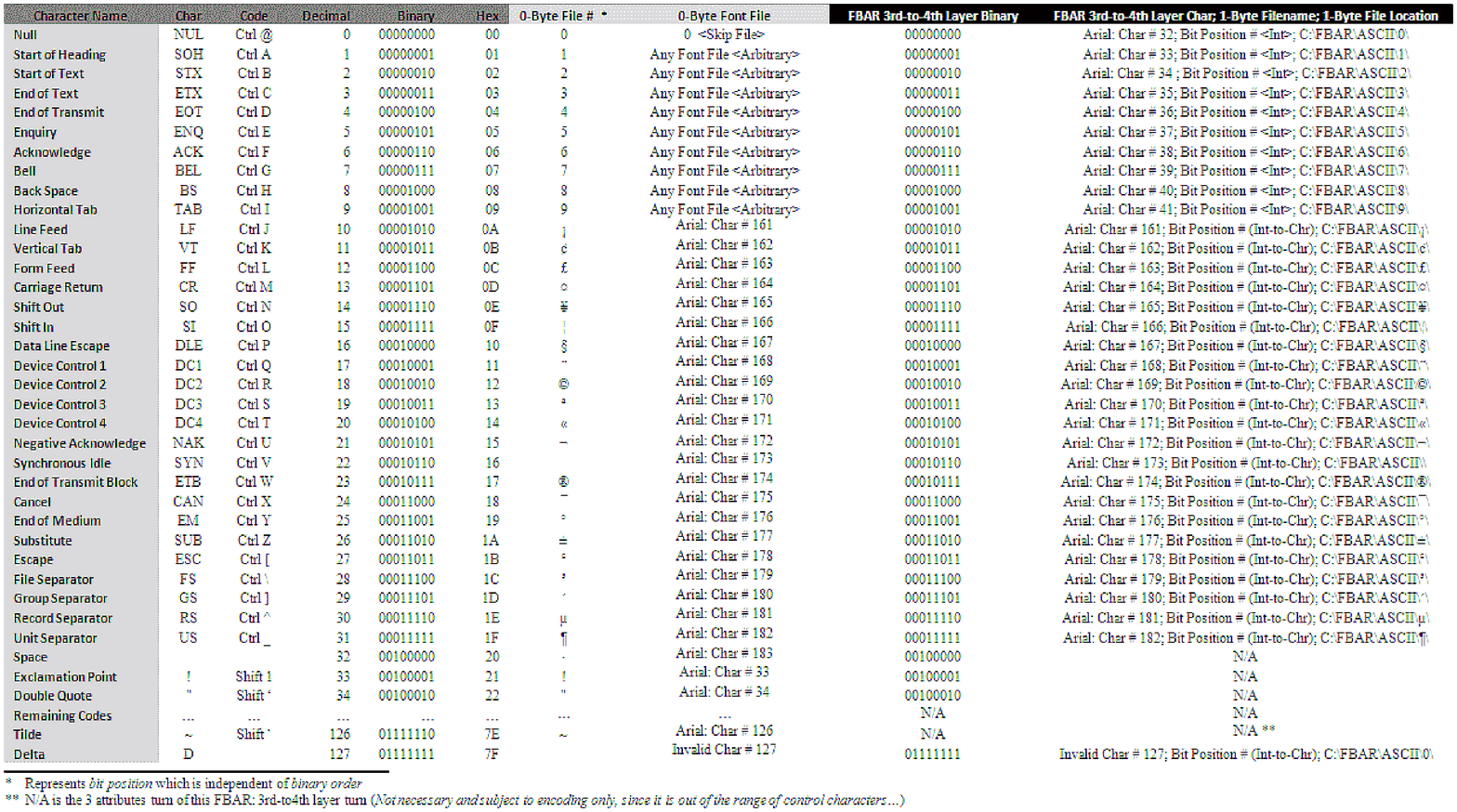}\\
\end{flushleft}
\noindent{\footnotesize{\textbf{Table.$\,$1.} This table is a new representation of ASCII table satisfying FBAR algorithmic requirements during layer-by-layer AND/OR 1-\emph{byte}, \emph{character-to-character} conversions. \\

\normalsize This avoids problems of encoding data in the final levels of compression which could end up in losing data, e.g., a 5 bit result cannot be converted to a 1-byte standard ASCII character, making it most difficult to decompress data in the later stages of the algorithm. For instance, in the ``\verb"My Car"'' case, string-to-binary gives out 48 bits = 6 bytes of information. For encoding purposes, we then apply further operations like ANDing and ORing, and as we experience from the 2nd layer of the compression phase, 24 ANDed bits and 24 ORed bits in parallel, after detecting pure binary pairs in one set like $\{00, 11\}$, and fuzzy binary pairs like $\{01, 10\}$ in the remaining set, the unbalanced length occurs when we reach conditions like line\,\#\,15 relative to the code sample presented in \S\,\ref{section3.4.2}. Since a substitute for pure combinations 00 and 11 could be encoded as 0 and 1 respectively, (\S\,\ref{section3.4.2}), then a fixed length is obtained, but for 01 and 10 further encoding is needed before string product conversion. Thus we are obligated to complete the length that has been changed into binaries of some random length, e.g., 2, 3, 4, 12, 14, ...  which are not in the proper range of fixed word lengths, converting them to printable string characters generate NULL characters that might or might not neighbor a proper string character indeed. The solution to this problem lies in adding at least a sequence of bits just like line\,\#\,24, to form a fixed length of bits, at a minimum of 8 bits, which is equivalent to one of the 1-byte standard ASCII printable characters. We have further elaborated on this in \S\,\ref{section3.4.2}. The progressing layers to the 2nd layer covering the pragmatics and analysis of the compression phase, commence as follows: \\

\noindent \emph{Exceptional character key assignments:}----------- We remove certain binary equivalent controls in the ASCII table during binary conversion to string of characters. For instance, the ANDed and ORed products of a simple two character input ``\texttt{PT}'' in the text field are 0000 and 1010, respectively. Once concatenated, results in ``00001010'' binary, as the 10th decimal which is equivalent to the line feed ``\texttt{LF}'' character with a ``\texttt{Ctrl\,J}'' code, or simply, a Not Applicable (\texttt{N/A}) control character in the ASCII table. In result, gives out a rectangular char after converting to string characters. To avoid this, a zero-byte directory of such \texttt{N/A} character controls substituting for \emph{special characters} is created, and a new numerical position of 1-\emph{byte} representing bitwise AND/OR standard conversion between characters is set. To apply exceptional conversions under these binary conditions, we code. Here goes a sample code in VB:

\scriptsize
\begin{flushleft}
\noindent---------------------------------------------------------------------------------------------------------------------------------\\
\texttt{0 \ Dim next\_char, result(0), bin(0) As String\\
1 \ Dim i As Integer\\
2 \ Dim ascii As Long\\
3 \ result(0) = ""  'assign a null or 0 string length\\
4 \ FOR i = 1 TO Len(bin0) + 18 STEP 8\\
5 \ next\_char = Mid\$(bin(0), i, 8)\\
6 \ ascii = BinaryToLong(next\_char)\\
7 \ IF next\_char = "00000000" THEN ' An exceptional conversion to ASCII code, e.g., \# 32 \\
8 \ result(0) = result(0) \& " "\\
9 \ ELSE\\
10       result(0) = result(0) \& Chr\$(ascii) 'don't miss with other characters \\
11        END IF\\
12    NEXT i}\\
\noindent---------------------------------------------------------------------------------------------------------------------------------\\
\textbf{Note}: The \texttt{LongToBinary} function converts Long value into a binary string needed for preliminary steps of ASCII conversions before string-to-binary. \\
\end{flushleft}
\normalsize
\noindent which denotes the use of an exceptional conversion method over 00000000 as a non-printable NULL into SPACE (ASCII character \#\,32). More IF statements for other characters follow the newly-applied rules in Table\,1. Hence, selecting and substituting these characters as an ASCII character control alternatives, becomes valid to our use (see Table.$\,$1). This approach could be classed as a permissible efficient model to encode a character in form of 1\emph{byte-to-}1\emph{byte} conversion rather than other alternatives which occupy more space, e.g., 1\emph{byte-to-}2\emph{bytes} $\equiv$ 1\emph{char-to-}2\emph{char} \emph{unicode} conversions \cite{24-Jukka_Korpela}. The two leftmost columns of Table 1 represent reliable conversions when stored in an identifiable zero-byte directory like the depicted folder icon labeled with ASCII in Fig.$\,$5. Moreover, these two columns in their rows as one studies never intersect with lower rows of an ASCII map i.e., decimal numbers between 32 and 127, and just a use for encoding the control character to a printable one in some separated location which does not share decimal due to 4-\emph{bit} + 4-\emph{bit} concatenation equivalent to a byte character of control type. The default location as specified is the root (\texttt{C:$\backslash$}) directory when bits are produced from the third layer to forth after AND/OR application. The range is of Integers (or Int) for 0 to 9, and the rest form the same characters borrowed from the lower-middle towards lowest ASCII map font characters. If the bit position in all of the rows indicates double digit values, we virtually convert Int-to-Char, meaning that, e.g., say we encode bit position \#\,10 to its equivalent from the lower ASCII map characters. The answer is Arial Char \#\,161. The choice of Font is accustomed to the need of programmer which attempts to distinguish invalid chars (e.g., little rectangles) from nulls and printable ones neighboring those without any output, e.g.  00000000, 00001010 and etc. (most significant chars). The bit position, denotes the bit that ends (or preferably begins in sign of an 8-\emph{bit flag sequence}) with a byte of an ASCII control character. -----------

\subsection{A zero-byte batch file solution to current data compression standards}
\label{section4.1}
In virtue of having a lossless data decompression in form of binary with excluded fuzzy and quantum noise indicators, explicated in \S\,\ref{section4}, all characters fallen into the conversive sequences of bits with either indicator, ``$\slash$" or ``$\backslash$", could be assumed to have at least a full state or 1$:$1 quantum noise inclusion for ``01" or ``10" paired bit conditions. Instead of thinking within the norms of $\frac{1}{2}$ a bit, one could conceive the concept in all halved bits as added up units of data to give 1, or \vspace{-1mm}
\begin{equation}\label{13}
\frac{1}{2} \ bit  + \frac{1}{2} \ paired \ bit = 1 \ full \ state \ bit \,
\end{equation}

\long\def\symbolfootnote[#1]#2{\begingroup%
\def\thefootnote{\fnsymbol{footnote}}\footnote[#1]{#2}\endgroup}

\noindent thereby not losing information on the compressed data, $C(x)$, Shannon entropy scales in aim of conserving the concept of lossless data compression vs. lossy conditions. To do this, we currently used the notion of zero-byte files, integrally. The zero-byte file as a computer file contains no data and during its creation (considering filename and other file properties) might occupy some weight especially when batched in form of a set of OS-Shell commands either in DOS, or in a more resilient and rich shell language, UNIX shell, or compressed via e.g., ZIP ranging from 1-byte to 1-kilobyte. We examined this for a single digit and a double digit zero-byte filename, virtually having a size of 100 and 102 bytes respectively. \symbolfootnote[1]{ These sizes represent the actual file size after ZIP compression much lesser than the files' physical HDD cluster (\emph{a group of disk sectors}) occupation, where the size on disk for that zero-byte or any type file will reflect the entire cluster as being used.} A zero-byte file is usually an OS error related object during incomplete file download sessions, prematured program interruptions and etc.~\cite{15-Wiki}. In this case, we purposely create a set of zero-byte (zero-length) files, of course, as not hundreds and thousands due to a potential memory crunch occurrence. This is merely to quantify our quantum noise matrix forming our decompression's input bit values, including their bit position trough program code quantum functions. Furthermore, to quantify, we allow batch file creation after a limited set of zero-byte file generation from the FBAR program. The batch file(s) contain(s) the numbers of quantum bit positions including their polarities, and for double, triple and ... digits, we consider the equivalent character in the standard ASCII table, e.g., the zero-byte filename 126 as a decimal number out of 128 ASCII characters of that table assigned in the batch file is substituted for the $\sim$ sign. Of course, the 8-bit ASCII extended characters and control characters from decimal \#\,128 to 255, and \#\,0 to 31, respectively are considered with relevant character control encodings expressed in~\S\,\ref{section3.4.3}.

For the FBAR data compression phase preparing characters for a decompression phase, we strictly avoid \emph{unicode} ASCII conversions since the disadvantage lies in occupying larger space between planes of bytes sequences due to a \emph{variable-length character encoding} characteristic \cite{24-Jukka_Korpela}. The Unicode code-points are logically divided between 17 planes, logarithmic to the 4 possible byte combinations, representing a word size of 16 bits, giving out 65,536 possible values ($=2^{2^{2^2}}$) code points~\cite{23-Weisstein}.

Ergo, within the FBAR conversive components, the concept of characters of 1-byte to 4-byte encoding characters is casted away. The main focus is on those points that maintain 1-byte-to-1-byte conversions only (0-127 ASCII characters) whilst extended characters of, e.g., Asian languages conserved in their context~\cite{25-Pike_Tompson}.

Henceforth, for maintaining better entropic releases and lower bit/character percentages, resulting higher data compression ratios, e.g., values $\geq 2$$:$$1$, the 1-byte character substitutions are conducted in the FBAR algorithm.

Filename 0 is always set to NULL, and no fuzzy quantum indications nor otherwise in position. During decompression, this is skipped as a header, reading next as the first fuzzy quantum bit position. For a binary code of some sort, e.g., the \emph{null byte} with 00000000 binary representation, as a control character `\texttt{Ctrl\,@}', a special character substitute is suggested during FBAR Char-to-Binary and vice versa AND/OR conversions (explained earlier in~\S\,\ref{section3.4.3}). We then iterate via a For Loop as the next file matrix block, reflecting the next section of the 4th layer string (if too big in content) into recurring filename decimals.

A full fuzzy quantum state representative is to be used for the decompression sequence of information stored with `/' for right polarity rising from Low-to-High state logic and values for left polarity falling from High-to-Low state logic, for a 24-bit sequence, in terms of, e.g., QLAND01, QLAND06, QLAND19 and ..., parallel to, e.g., QRAND06, QRAND11, QRAND22 and ..., exclusively, ``exact'' to the concluded and stored sequence of characters of the compression's 4th layer. The QLOR and QROR have a similar approach parallel to QLAND and QLOR. Accordingly, We assumed QAND as the ANDed values in the $C$'s 4th layer just for a fuzzy quantum indication, and QOR as the ORed values parallel to AND, in the same layer for a fuzzy quantum indication. The appended ``$L$'' and ``$R$'' in the filename in their respective file directory (or \emph{quantum memory address} instead) represent ``left" and ``right" polarities, respectively.

\section{Decompression layers}
\label{section4}
It is at this stage significant to examine the compressed data in packets of sink data when user commands to decompress the resulted data, aiming for data integrity upon original values. Hence, the decompression procedure begins with the last compression layer i.e., the bottom of the compressor's 4th layer. This is recalled by Fig.$\,$7\,\emph{b}) establishing a multiplexer's usage for two fqubit registers. However, the technical definitions in virtue of exclusion operators differ from the definitions represented within the previous data compression sections.

\subsection{4th layer: Exclusion from noise with fuzzy logic indicators}
\label{section4.1}

\normalsize The current FBAR as FQAR supports the full fuzzy quantum states illustrated in~\S\,\ref{section3.4}, i.e., the IN's and OUT's as it progresses in specific quantum state comparisons during $C$ thereby $C'$, an ASCII map of the equivalent character to the batched memory address representing ceratin character encodings as substituters for a set of combined 0 and 1 dot properties. Nevertheless, the FQAR seclusive proposal (\S\,\ref{section3.4.1}) remains intact to the constituents of our concept, supplying relevant quantum components for our $n$-level fuzzy quantum mechanical system rather than just the well-known two-level quantum system.

Therefore, for excluding noise, one must conceive the reality of which congruent multi-leveled bit data support in form of dots from the atomic lattices in position, as necessary components for ultimate compression, here lossless decompression. The noise descriptive, of course, are dots carried by at least 2-fqubit waves, the extraction is evident to the revolution of periods that have been occupied in the atomic space. These fragmented data are in result defragmented by the multiplexer via \emph{select}, \emph{a noise excluder of which signals are dissected for putting together dots via reverse projection}. Dots have distances of precise $\lambda /n_i$ between each other, denoting one of their properties, either of address type, position, polarity or logic state per half a $2\pi$ revolution, all with a bitrate (\emph{intensity}) with a set of reverse numerals to ``indicators'' symmetric to the other half of any other revolution; hence the dots could be transferred to that side of the carrier (wave). This reverse projection reads the 4-\emph{compressed bit properties} explained in \S\,\ref{section3.5}, hence the selection must emit four groups no matter how many trials attempted, four sets to be ensured for data integrity. This is dependant upon how we store the whole information between singularities of coupled co-variant signals (Fig.$\,$7) with precise clocking for synchronicity in preserving data when spatial locality of reference reserved for the group of dots on lattice sites. The preservation of data when decompressed, must conform to the limits of reverse exclusions not losing content and thus data integrity from the highest level of compression to lower ones as follows:

\subsection{3rd layer: ANDed pure pairwise-to-OR exclusion technique}
\label{section4.2}

\normalsize The tricky part to the reverse exclusions techniques of the FBAR algorithm is this layer of $C'$ integration. The ANDed pure pairwise or, the detected pure pairwise Boolean set in \S\,\ref{section4.3}, is now excluded to obtain decompressed values close to the notion of $C$. After excluding relative to ORed values, then we once again have two sets of binary sequence of: 1- ORed values with a closure binary point, a bit value = \textbf{0}, and 2- ANDed values with a closure binary point, a bit value = \textbf{1}. These two values will act as a flag for our algorithm executing an exclusion code after a fixed word length conversion (after blocks of 8, 16, 32, 64 or 128 bits, considering nibbles with 4-bit length representing a null or space) from the 2nd layer of the compression phase. The exclusion code must by flag \textbf{0} give out a sequence of ORed, and flag \textbf{1} a sequence of ANDed values side-by-side confirming the fuzzy state of our FBAR algorithm.

\subsection{2nd layer: AND/OR pairwise pattern match with iterative search technique}
\label{section4.3}

In content, the current layer aims at pre-ORed and pre-ANDed bits through pattern matching i.e., original data as once inputted in this case, ``\texttt{Philip Baback}", but at a binary level. After achieving this, binary-to-string is subject to 1st layer decompression in \S\,\ref{section4.4}. There is no miss-matched bit individuals encountered in this technique, hence the concept of lossless data reserved. \\

\noindent 0000 0000 0000 0010 0000 0100 \textbf{1} 00 0000 0000 0000 0000 0001 0001 \textbf{1}\\
$\downarrow\downarrow\downarrow\uparrow \ \downarrow\uparrow\uparrow\downarrow \ \downarrow\uparrow\uparrow\downarrow \ \downarrow\uparrow\uparrow\downarrow \ \downarrow\uparrow\uparrow\downarrow \ \downarrow\uparrow\uparrow\downarrow$ $ \downarrow \ \downarrow\downarrow  \ \downarrow\downarrow\downarrow\uparrow \ \downarrow\uparrow\uparrow\downarrow \ \downarrow\uparrow\uparrow\uparrow \ \downarrow\uparrow\uparrow\downarrow \ \downarrow\uparrow\uparrow\downarrow \ \downarrow\uparrow\uparrow\downarrow \ \downarrow$\\
\noindent 1100 1110 1111 1110 1111 1100 \textbf{0} 01 1001 1101 1101 1101 1101 1111 \textbf{0} \\

\noindent The pattern match indicator in both `\texttt{P}' and `\texttt{B}' cases are $\downarrow\downarrow\downarrow\uparrow$, which denotes that these capital letters have the same pattern behavior contrasted to the lower capital letters with indicator $\downarrow\uparrow\uparrow\downarrow$.

\scriptsize
\begin{tabbing}
  \emph{Type no.} \= \emph{Polarity set} \ \ \ \ \  \= \emph{Implies to \ \ \ \ \ \ \ \ \ \ \ } \= 1-\emph{bit flag per nibble}\,; $\beta$ \ \ \ \ \ \ \ \ \= \emph{fqubit type\dots}\\
  0 \> $\downarrow\uparrow\uparrow\downarrow$ \> most characters \> $f_{0}$=1-bit ; $\beta_0$ = 0.125 \> $]0,\frac{1}{2n}[ \equiv ]0, 0.125[$\\
  1 \> $\downarrow\downarrow\downarrow\uparrow$ \> letters \> $f_{1}$=1-bit ; $\beta_1$ = 0.125 \> $]0,\frac{1}{2n}[  \equiv ]0, 0.125[$ \\
  2 \> $\downarrow\uparrow\downarrow\downarrow$ \>  letters \> $f_{2}$=1-bit ; $\beta_2$ = 0.125 \> $]0,\frac{1}{2n}[\equiv ]0, 0.125[$ \\
  3 \> $\downarrow\uparrow\downarrow\uparrow$ \>  letters \> $f_{3}$=1-bit ; $\beta_3$ = 0.125 \> $]0,\frac{1}{2n}[\equiv ]0, 0.125[$ \\
  4 \> $\downarrow\uparrow\uparrow\uparrow$ \> letters \> $f_{4}$=1-bit ; $\beta_4$ = 0.125 \> $]0,\frac{1}{2n}[\equiv ]0, 0.125[$ \\
  5 \> $\downarrow\downarrow\uparrow\uparrow$ \> few letters \> $f_{5}$=1-bit ; $\beta_5$ = 0.125 \> $]0,\frac{1}{2n}[ \equiv ]0, 0.125[$\\
  6 \> $\downarrow\downarrow\downarrow\uparrow$, $\downarrow\uparrow\uparrow\downarrow$, ... \> dual characters \> $f_{6}$=1-bit ; $\beta_5\in$ [0.125, 0.75] \> $]0,\frac{1}{2n}[\equiv ]0, 0.75[$\\
  7 \> $\searrow$\>  all 2-bit binary 10  \> $f_{7}$=1-bit ; $\beta_7$ = 0.5\> $]0,\frac{1}{2n}[\equiv ]0, 0.125[$  \\
  8 \> $\nearrow$  \> all 2-bit binary 01  \> $f_{8}$=1-bit ; $\beta_8$ = 0.5 \> $]0,\frac{1}{2n}[\equiv ]0, 0.125[$ \\
\end{tabbing}
\vspace{-2mm}
\noindent{\footnotesize{\textbf{Table.$\,$2.} This table is an fqubit representation customized on memory and process flags as a layer-by-layer AND/OR \emph{1/2 byte-to-1/2 byte} pairwise polarity detection(s), satisfying FBAR algorithmic fuzzy AND/OR entropy plus bit frequency cover requirements. \\

\normalsize \noindent Moreover, some other lower capital frames carry indicator $\downarrow\uparrow\downarrow\downarrow$ like letters `\texttt{d}, \texttt{e}', and $\downarrow\uparrow\downarrow\uparrow$ like letter `\texttt{f}'  , and  $\downarrow\uparrow\uparrow\uparrow$ like letter `\texttt{j}', and some fall into two or more categories like letters `\texttt{g}, \texttt{k}' as \emph{dual polar characters}. Based on this type of classification, Table 2 ascertains data representing indicators in terms of \emph{type}\,\# 0, 1, 2, 3, 4, 5, 6, 7 or 8 which could be stored adjacent to 1-\emph{bit flag} reference to a particular type of polarity into a bit-field pack memory structure otherwise, for an FQAR quantum state, into a bit-field pack signal structure: to be in-between 2-\emph{encoded bits} i.e. a quantized signal (continuous time, discrete values) encoding from an infinitely possible two-level bits (from a pool of \emph{n}-values) to high level (e.g., $\pi = 3.1415...$ for 1) and low level logic (e.g., $e=2.7183...$ for 0) in computer circuitry~\cite{35-Tocci}. In this table, $\beta$ denotes a \emph{bit frequency to be covered between poles}, ideally for decompression, e.g., to reconstruct 1-\emph{char}, a classical bit frequency to be covered for $\downarrow\uparrow\uparrow\downarrow$ would be $f_{0}/\{(1_{bit_0 \vee} + 1_{bit_0 \wedge}) + (1_{bit_1 \vee} + 1_{bit_1 \wedge}) + \ldots + (1_{bit_7 \vee} + 1_{bit_7 \wedge}) \} = 1/8 = 0.125$. The one row reflecting dual characters gives $1/8=0.125$ to $6/8=0.75$ frequency due to the nibble in which may apply to the 6 previous rows, i.e., polar combinations. Perceivably, the last two rows have a $\beta= 0.5$, general to fuzzy binary type recognizable from \S\,\ref{section3}. To distinguish one polarity type from another, we address the orientation of polarities per character in a set of \emph{virtually paired equalities }detected by polarity identifiers, turning them into \emph{really paired inequalities} as follows:

\subsubsection{Odd orientation of bit polarity per character}
\label{section4.4}

As we computed the conversions of 1st layer towards the 4th layer, we realized an \emph{odd} versus \emph{even} orientation of most significant bit (\emph{left-most bit}) polarity prior to uppercase alphabet detection compared to lowercase alphabet character illustrated in the 2nd layer of decompression layers, \S\,\ref{section4.3}. Benefiting from Conversion Rel.\,(\ref{9}), as we perceive, the characters' binary states are exactly equal, giving no probability after being ANDed and ORed simultaneously. Let $\eta$ represent an English alphabet odd orientation, then after $\wedge\vee$ conversion to binary \emph{bin}, we deduce
\[\forall (X_{\eta}\textsl{X}_{\eta+1}, \textsl{x}_{\eta}\textsl{x}_{\eta+1}) \in \mathcal{A}  {\left( {X,\textsl{x}} \right)}\stackrel{\wedge\vee}{\longrightarrow}f(\emph{bin}) \ , \ \exists \, \eta \rightarrow i \in 2\mathbb{Z}^* \ , \ j \in 2\mathbb{Z}^*\!+ 1 ; \]
\[\mathcal{A}  {\left( {X,\textsl{x}} \right)} = \{(AB, ab), (EF, ef), (IJ, ij), (MN, mn), (QR, qr), (UV, uv), (YZ, yz) \} \]
\[\mathrm{iff} \ \mathcal{O}(\uparrow,\downarrow)\longrightarrow \mathcal{O}(\uparrow\uparrow\in\nearrow\searrow)= \mathcal{O}(\uparrow\downarrow, \downarrow \uparrow, \downarrow\downarrow \in\nearrow\searrow) \]
\[\therefore (A=B, \ a=b), \ (E=F, \ e=f), \ (I=J, \ i=j), \ (M=N, \ m=n), \  \]
\[(Q=R, \ q=r),\ (U=V, \ u=v), \ (Y=Z, \ y=z)\]
\noindent thus for $C_{\wedge\vee}$
\[ \therefore (X_{2i+1}=\textsl{X}_{2i+2}, \textsl{x}_{2i+1}=\textsl{x}_{2i+2})_{bin} \  \mathrm{and} \ (X_{2j+1}\neq\textsl{X}_{2j+2}, \textsl{x}_{2j+1}\neq\textsl{x}_{2j+2})_{bin} \ , \]
\noindent $\mathrm{if} \ \mathcal{O}(\uparrow \neq \downarrow)$ for $C'_{\wedge\vee}$, then
\begin{equation} \label{12}
 \therefore (X_\eta\neq \textsl{X}_{\eta+1}, \textsl{x}_{\eta}\neq\textsl{x}_{\eta+1})_{chr} \ , \ (X_\eta\neq \textsl{X}_{\eta+1},\textsl{x}_{\eta}\neq\textsl{x}_{\eta+1})_{bin} \, . \
\end{equation}

\noindent where $\mathcal{O}$ is the function of Bottom-to-Top and Top-to-Bottom orientation identifier, \emph{if and only if} not detected, null or dictating a codal detection of becoming equal i.e. $\uparrow = \downarrow$, then the odd orientation relative to their symmetry is always \emph{true}. Otherwise, when  $\uparrow \neq \downarrow$, the odd orientation is \emph{false} and thus detection of such binary recursions after AND and OR operations is in position. So, in general, we derive $bin_1=bin_2, bin_5=bin_6, \ldots$ showing that out of 26 letters, we have done a 2:1 compression with hidden polarities, i.e. 14 for $\eta$'s odd orientation, 12 for $\eta$'s even orientation, thus giving out for capitals with lower type case $14\times 2 = 28$ compressed characters. To satisfy this identifier requirement, we singularly examined characters $a$ and $b$, and made relevant comparisons to refine program statements for distinguishing polarities in case of mismatch between such aligned characters in their binary state. This is given by the following pseudocode:

\scriptsize
\begin{flushleft}
\noindent---------------------------------------------------------------------------------------------------------------------------------\\
0 \texttt{START FBAR/FQAR COMPRESSION \\
1 IF strChar = Chr\$(97) OR \emph{any Char that follows the same flag as} "a" THEN  \\
2 SET ATTRIB TO 0 = 1-\emph{bit}  'allocate 1 bit as flag 0 denoting DOWN-UP-UP-DOWN \\
3 STORE 1-\emph{bit to memory} 'we raise this flag when needed for decompression \\
4 END IF\\
5 IF Str(Char) = Chr\$(98) OR \emph{any Char that follows the same flag as} "b" THEN \\
6 SET ATTRIB TO 4 = 1-\emph{bit} 'allocate 1 bit as flag 4 denoting DOWN-UP-UP-UP \\
7 STORE 1-\emph{bit to memory} 'we raise this flag when needed for decompression \\
8 END IF\\
9 ... 'we code other similar IF statements \\
10 OUTPUT\,\#1:Char "a" Polarity Product = 0x\emph{pppppppp} 'a classical polarity address \\
11 OUTPUT\,\#2:Char "b" Polarity Product = 0x\emph{pppppppp} \\
12 ...\\
13 OUTPUT\,\#\emph{n}:Char <\emph{x}> Polarity Product = 0x\emph{pppppppp}\\
14 HALT FBAR/FQAR COMPRESSION } \\
\noindent---------------------------------------------------------------------------------------------------------------------------------\\
\end{flushleft}
\normalsize

\vspace{16mm}
\bigskip
\vspace{40mm}
\begin{flushleft} \hspace{-1mm}
\includegraphics[width=3.35mm, viewport= 0 0 10 10]{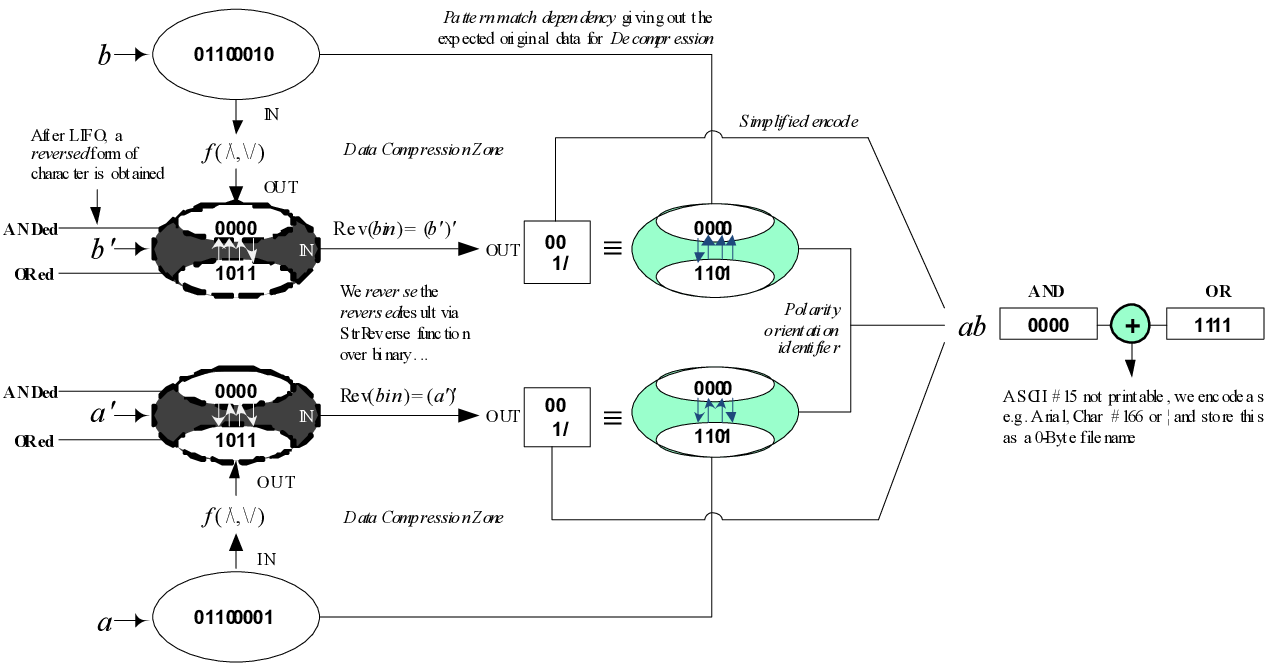}\\
\end{flushleft}

\noindent{\footnotesize{\textbf{Fig.$\,$8.} A schematic view of how data decompression occurs for two simple characters `$a$' and `$b$' in the FBAR/FQAR algorithm. The core of the algorithm is validated via its polarity orientation ability next to none if not considered for retrieving data, or i.e., initial data reconstruction from the point of compressor's input to the point of decompression's output. \\

\normalsize
\noindent This example is illustrated in Fig.$\,$8, and gives a simplified encoded solution to the IV-layer product at a decompression state, commencing with storing relevant polarity information into minimum size of space during IV-layer compression (the zero-byte file solution). Now, if the character during the layer-by-layer conversions, carries a QLOR, QROR, QLAND and/or QRAND polarities, as if address \#\,\texttt{0x1}, \texttt{0x2}, \texttt{0x3}, \texttt{0x4} with type 7 or 8 from Table\,2, we then code for decompression by displaying results in a text field:

\scriptsize
\begin{flushleft}
\noindent---------------------------------------------------------------------------------------------------------------------------------\\
0 \texttt{START FBAR/FQAR DECOMPRESSION \\
1 PUBLIC FUNCTION InsertString ( ByVal\, vDest \, As\, Variant , ByVal\, vSource \, As \, Variant \_ \\
2 \ \ \ \ \  \ \ \ \ \ \ \ \ \ \ \ \ \ \  \ \ \ \ \ \ \ \ \ \ \  , Optional ByVal vInsertPosition As Variant) As Variant \_ \\
3 ... 'we code other special functions like Replace and Cat (catenation) for polarity flags
4 PRIVATE DECLARE SUB GetDWord Lib "FBAR\_FQAR.dll" Alias \_ \\
5    "GetMem" (ByRef inSrc As Any, ByRef inDst As Long) \\
6 PUBLIC FUNCTION DeRef(ByVal inPtr As Long) As Long \\
7    IF (inPtr) THEN CALL GetDWord(ByVal inPtr, DeRef) \\
8 END FUNCTION \\
9 ...} \\
\noindent---------------------------------------------------------------------------------------------------------------------------------\\
10 \texttt{'to read from memory, regA contains \emph{bit pos.}and \emph{bin}, regB for \emph{bit pol.}and \emph{bit addr.} \\
11 Dim pol As Long\\
12 Dim regA As Long, regB As Long \\
13 Dim myPtr As Long 'for atomic selector or variable pointer varPtr\\
14 SELECT CASE pol \\
15 'next lines infer to fuzzy quantum orientations\\
16 CASE 7 to 8 'this is QLOR, QROR, QLAND, QRAND \emph{pol.flag no.}\\
17 myPtr = varPtr(regA) 'get a pointer to memory variable that allocated pos for a char\,if\,any \\
18 'dereference prints \emph{bit pos.no.}and AND-OR \emph{bin} = $\{$2,\,4,\,0000,\,1111$\}$ or a list of 2-positions  \\
19 Debug.Print DeRef(myPtr) \\
20 myPtr = varPtr(regB) 'get a pointer to memory variable that allocated pol for a char\,if\,any \\
21 Debug.Print DeRef(myPtr)  'de-reference prints, \emph{bit pol.\,no.}and \emph{bit addr.}=$\{$7,\,0x2$\}$ which \\ 22 'means QROR only and the stored ANDed bin is 0000, ORed bin is 1111 \\
23 ...\\
24 Replace("11 11", "01", 2) = "101 11" '01 is the equivalent of $/$ for QROR \\
25 'we increment \emph{pos. no.} by 1 since the 1st replacement is over \\
26 Replace("101 11", "01", 4+1) = "101 101" \\
27 'next lines infer to classical orientations \\
28 CASE 0 to 6  \\
29 myPtr = varPtr(regB) 'get a pointer to memory variable that allocated pol for char if any \\
30 Debug.Print DeRef(myPtr)  'de-reference prints, \emph{bit pol.no.}and \emph{bit addr.}= $\{$0,\,4,\,0x0$\}$\\
31 Replace("00101  00101", "01", 1) = "01 0010111" 'we now apply flag 0 binary replacements \\
32 'we first expand our nibbles to bytes based on pure pairs \\
33 Replace("\textbf{00}101 \textbf{00}101", "00", 1) = "\textbf{000}101 \textbf{00}101" \\
34 Replace("\textbf{000}101 \textbf{00}101", "00", 2+1) = "\textbf{0000}101 \textbf{00}101" \\
35 ' now a nibble for AND is formed; since we encountered a pure 1 resultant; restart for OR  \\
36 Replace("\textbf{0000}101 \textbf{00}10111", "11", 1) = "\textbf{0000}1101 \textbf{00}101" \\
37 'now a nibble for OR is formed; now start forming a nibble for AND \\
38 ...\\
39 Replace("\textbf{0000}1101 \textbf{0000}101", "11", 1) = "\textbf{0000}1101 \textbf{0000}1101"\\
40 'now we have a sequence propped for classical polarities, \\
41 'we now apply flag 0 binary for a nibble-to-nibble concatenation according to \emph{pol.no.}turn \\
42 Cat("\textbf{0000}1101", 1, 5) = "\textbf{0}1 000 101" \\
43 Cat("\textbf{0}1 000 101", 5+1, 1+1) = "\textbf{0}11\textbf{0} 00 01" \\
44 Cat("\textbf{0}11\textbf{0} 00 01", 6+1, 2+1) = "\textbf{0}11\textbf{0}0\textbf{0} 0 1" \\
45 Cat("\textbf{0}11\textbf{0}0\textbf{0} 0 1", 3+1, 7+1) = "\textbf{0}11\textbf{0}0\textbf{0}\textbf{0}1" \\
46 'we now apply flag 4 binary for the second bin sequence with a similar approach \\
47 Cat("\textbf{0000}1101", 8, 1, 5) = "\textbf{0}1 000 101" \\
48 ...\\
49 Cat("\textbf{0}11\textbf{0}0\textbf{0} 0 1", 7+1, 3+1) = "\textbf{0}11\textbf{0}0\textbf{0}1\textbf{0}" \\
50 END SELECT \\
51 ...}\\
\noindent---------------------------------------------------------------------------------------------------------------------------------\\
\end{flushleft}
\normalsize
\noindent As a suppositional condition, we read the contents of an fqubit memory by \emph{de-referencing} data (e.g., lines\,\#\,21 and 30). In line \#\,17, we assumed address \texttt{0x0} is a null pointer, no attempt to access it and just noting that replacements or insertions of bits must occur from the leftmost bit available. This is where a successful decompression is evaluated for data integrity, thus data loss for FBAR/FQAR entropy is examined addressing statistical verification between characters and bits of original data.

\subsection{1st layer: Binary-to-string conversion}

This layer is easy to acquire for bits unless definitive in terms of fqubits. However, since the concept of dot properties has already been explained in the previous layers, we assume to have all data available in this layer, leaving the system for good after a successful binary-to-string conversion, thereby confirming that the present transformed data is in alignment with the previous layer statistics. The output for this is obvious and continues from line\,\#\,51 in the previous pseudocode:
\scriptsize
\begin{flushleft}
\noindent---------------------------------------------------------------------------------------------------------------------------------\\
\texttt{51 ...\\
52 Dim bin As String 'casting for the sole purpose of printing characters\\
53 Dim result As String\\
54 Dim i As Integer\\
55 Dim next\_char As String\\
56 Dim ascii As Long \\
57 'the binary of character <\emph{whatever}> after identifying its particular indicator \\
58 bin = "\textbf{0}11\textbf{0}0\textbf{0}\textbf{0}1" + "\textbf{0}11\textbf{0}0\textbf{0}1\textbf{0}" \\
59 For i = 1 To Len(Bin) + 18 Step 8\\
60        next\_char = Mid\$(Bin, i, 8)\\
61        ascii = BinaryToLong(next\_char)\\
62        result = result \& Chr\$(ascii)\\
63    Next i\\
64 OUTPUT: "ab"\\
65 HALT FBAR/FQAR DECOMPRESSION}\\
\noindent---------------------------------------------------------------------------------------------------------------------------------\\
\end{flushleft}

\normalsize
\noindent By this in account, the FBAR system turning into FQAR, led us to come up with relevant conjectures in computing the odds of our data transformation between servable spaces of binary, projected in form of dots and vice versa, by their extremely quantized signals. Moreover, the system is presumed to have a time interval passive to a constant refreshing loop phase of duals (\emph{bits of dual logic}) as a normalization procedure between fqubit registers, proposed in \S\,\ref{section3.4.1}, in form of ``bi-bottom lattice sites" relative to their top site. This leads us to formulate the principles of the implementation into two theorems with their cyclic hypotheses which reiterate in plausibility, once the grounds of our seclusive proposal evaluated with expected outcomes.

\section{Algorithm's summary in form of IV-layer hypotheses}
\label{section5}
\normalsize One aspect coming about on these elementary components of proof, i.e., the algorithm's IV-layers, is its cyclic binary space converted into other forms of information plus, every layer's output. We recall in summary, the overall properties of the FBAR design based on its implementation asserting the following theorems under the generalization of a IV-layer data conversions' proof: \\

\noindent \textbf{Theorem 1. }----------- \emph{Let every layer output be a substantial proof based on compression $C$ and relevant conversions of data. Its memory scope in the field of scalars stores a quantitative scale of data, assuming by continued fractions of 10 for $C_i$, whilst quality of data preserved in terms of its text in the world of binary, fuzzy and quantum subspaces, performing compression. Then $\forall C_i \in \ell\stackrel{P}{\longrightarrow}\ell' \stackrel{P}{\longrightarrow}\ell''$, it is evident the differential form}
\vspace{-1mm}
\[ C_A =\frac{ \int \ldots \int_{\max(bit_{\mathrm{in}})}^{\min(bit_{\mathrm{in}})} C(C_1, C_2, C_3\ldots,C_n)10^{-i}\mathrm{d}C_i}{\int \ldots \int t \,\mathrm{d} t_i} = \left|\frac{C^{i+1}}{(10 \, t)^{i+1}}\right| \]

\noindent \emph{as compressed bitrate areas under curve which hold good in a closed-like subspace of Hilbert type at a deeper level, permitting a perpendicular projection $P$ map as if ``dropping the altitude'' of a binary triangle onto the planes of binary from one compression layer to another, generates conditions of Eqs.\,(\ref{2}-\ref{8} and \ref{10}) within the Hilbert space norm. -----------} \vspace{2mm}

\noindent One could thus deduce the triangular form in terms of a co-product \vspace{-2mm}
\[\xymatrix{
& \ell'' &\\
\ell \ar[r]_{C_i}\ar[ur]^{P} & \ell\coprod \ell'\ar@{-->}[u]^{P^2} & \ell'\ar[l]^{C_{i+1}}\ar[ul]_{P} }\]\vspace{-3mm}

\noindent which merely shows the mappable topological spaces dedicated to layers $\ell , \ell'$ and $\ell''$, correspondingly. These subspaces of Theorem 1, lead us to the crude representation of Eq.\,(\ref{10}), and are based on the relativity (its \emph{special theory}), qualitative and quantitative factors for compression such as: \emph{magnitude}, \emph{function} and \emph{dimensions} of the \emph{co-involved time-varying fuzzy-binary-quantum field} in space-time geometry. Hence for decompression \\

\noindent \textbf{Theorem 2. }-----------  \emph{Let the n-dimensional spatial involvement between subspaces of $\mathcal{H}$ appear always dual representing some co-involved time-varying fuzzy-binary-quantum field living in $\mathcal{H}^\textbf{*}$. Therefore, the base involvement prior to any fuzzy-quantum co-involvements in $\mathcal{H}^\textbf{*}$, is of course ``binary" for every $x$ bit input within these subspaces; at time $t$, constructs a rising space of $H_i$ with layers, or, $H_{\ell\ell'\ell'',t}$,  proportional to an increasing length function $\lambda(x)$ of ANDed and ORed pairs $(x_1, x_2)$ to a layer in form of a direct sum }
\vspace{0mm}
\[ C'_A \stackrel {\mathrm{extract}}{\longleftarrow_{\sum_i bit_i}} \mathcal{H_{\ell\ell'\ell''}}\int \mathrm{d}t = \bigoplus_{\lambda \in \ell\ell'\ell''} \mathcal{H}_\lambda =  \ \ \ \ \ \ \ \ \ \ \ \ \ \ \ \ \ \ \ \ \ \ \ \ \ \ \ \ \ \ \ \ \ \ \ \ \ \ \ \ \ \ \ \ \ \ \ \] \vspace{-3mm}
\[\! \langle \lambda(x_1), \lambda(x_2) \rangle\mathcal{H_{\ell}}^\textbf{*}+ \langle \lambda(x_1), \lambda(x_2) \rangle \mathcal{H_{\ell'}}^\textbf{*} + \langle \lambda(x_1), \lambda(x_2) \rangle \mathcal{H_{\ell''}}^\textbf{*}= \ \ \ \ \ \ \ \ \ \ \ \ \ \ \ \ \ \  \] \vspace{-3mm}
\[\!\!\!\!\!\!\!\!\!\!\!\!\!\!\!\!\!\!\!\!\!\!\!\!\!\sum\limits_i { \parallel \lambda({x_i }) \parallel} ^2 < \infty\ \mathrm{is \ complete \ and \ finite}\, ,  \ \ \ \ \ \ \ \ \ \ \ \ \ \ \ \ \ \ \ \ \ \ \ \ \ \ \ \ \ \ \ \ \] \vspace{-2mm}

\noindent \emph{with different ranks given in \S\,\ref{section2} during decompression, the notation $\stackrel {\mathrm{extract}}{\longleftarrow_{\sum_i bit_i}}$ is valid in the given context for extracting bit-by-bit of information to reconstruct data at the state of $C'$ onto a certain area $A$ of memory, whereby composite limits of extraction is in the scalar magnitude of either duals or pure states (for duals, see \S\,\ref{section3.4}). -----------} \\

In virtue of the \emph{ergodic hypothesis}, we currently apply and expand this hypothesis to such spaces that we established giving the following IV-layer hypotheses:\\

\noindent \textbf{Hypothesis 1. }----------- \emph{Over periods of a lossless data compression time, the time spent by the FBAR algorithm in some region of bit phase, the space of microstates' conversion with left and right AND/OR 00, 10, 11, 01 logic, relative to their initial binary input is proportional to the volume occupied by fuzzy and quantum bits of this region, i.e., all accessible microstates are equiprobable over an instant unit of time, computing AND/OR paired units of byte. This is now called an FQAR algorithm with fuzzy qubit property.
}-----------\\ \vspace{-2pt}

\noindent Therefore, an indicator mathematically representing more or less than a 1/2 paired bit to a 1 full state bit from Eq.\,(\ref{10}), promotes the above hypothesis to a state where \emph{invariant measures} preserved by function $\Delta\Delta\lambda$ for string length, and function for data content from compression $C$ to decompression $C'$ phases into this one\\

\noindent \textbf{Hypothesis 2. }----------- \emph{The FBAR IV-layer compression during its system evolution ``forgets'' its initial state i.e., the compression final output statistically $\neq$ initial output. Once we promote the fuzzy quantum inclusions into their exclusions on the decompression level, its four layers despite of encoded pattern complexity between one layer and another, store data for their logic state, address, position and polarity. It suffices to have a pattern respecting any polarity in a reiterative course between 0's, 1's and in-between states projecting fuzzy and quantum logic. Thus, the notion of data compression ratio $C_r \in [2$$:$$1, 2n$$:$$1]$, where $n\geq 1$ for a compression mapping ratio of $1$$:$$\texttt{len}(\texttt{bin})$ is always true, maintaining lossless entropy with a fractal behavior.} ----------- \\

The statistical geometry of this behavior on $C_r = 2n$$:$$1$ could be compared with the \emph{monitor-inside-a-monitor} effect from video feedback examples in \emph{chaos theory} \S\,1.1, Ref.~\cite{26-Peitgen et al.}. One could perceive the expansion of $n$$:$$1$ reflecting the length of \emph{$\texttt{bin}$}, or \emph{$\texttt{len}(\texttt{bin})$}, via its length of string on the decompression phase.

In fact, the infinitesimal fractal form from a \emph{string length mapping ratio} is of $1$$:$$\emph{\texttt{len}}(\emph{\texttt{str}})$, where $\emph{\texttt{len}}(\emph{\texttt{str}})\leq1$, exhibited on the compression phase for a long time evolution in the FBAR system when tried upon clusters and massive DBs. Thus, on the decompression phase performs $\emph{\texttt{len}}(\emph{\texttt{str}})\geq1$. Time for such transformations, poses itself as increments of $t$ during sequential compressions between nodes $\geq 2$ processors on some server's communication lines between source and sink DBs.

The communication lines could also be described as the OSI model (\emph{open system interconnection reference model}) seven layers from media bit physical layer to data application host layers and vice versa, established between $\min(n) = 2$-DBs. Ergo, a new concept of tracing data on those lines between DBs as \emph{transmittable buffer packets}, is notable in our future reports.

\section{The standardized IV-layer operation result}
\label{section6}
\vspace{6pt}
\noindent The current data compression assumptions from the previous section could be tested as facts and compared with based on one's approach within the context of entropic analysis (see, e.g., Refs.~\cite{{17-Shannon,{18-Shannon},{30-Shannon},{34-MaximumCompression}}}). For instance, the large text file described in the Statistical Distributions of English Text (containing the seven classic books with a 27-letter English alphabet) has a compression ratio of 36.3\% (original size = 5,086,936 bytes, compressed size = 1,846,919 bytes, using the Linux ``gzip'' program written in C language). This corresponds to a rate of 2.9 bits/character --- compared with the entropy rate of 2.3 bits/character predicted by Shannon. This loss of optimality is most likely due to the finite dictionary size, \S\,IV,~\cite{30-Shannon}. In the FBAR data compressor, however, we demonstrate that the input of a string sequence of English alphabets results in a 28-byte physical data, as a bin file stored onto a typical HDD. By taking into consideration the latter, a valid claim in having a temporary DB unit, convertible from a string type to binary and vice versa, relative to the nature of the FBAR algorithm comes to our attention. The AND/OR compression product would turn out to be the following based on the alphabetical odd orientation discussed in \S\,\ref{section4.3}:




We have proceeded with all compression and decompression layers inclusive of the 26$\times$2-English letters character set (excluding the space character) as a standardized approach to prove our hypotheses, \emph{true}. To fulfil this, we applied AND/OR, fuzzy and quantum operations, correspondingly. Accordingly, after a set of conversions and projections with relevant module design for compression, we expected from the FBAR compression model to eventually give out for any number of characters, this
%
%


\vspace{-3mm}
\begin{equation}\label{15}
C_{r} = \frac{\mathrm{Initial \ Data \ Size}}{\mathrm{Compressed \ Size}} = \frac{\texttt{\emph{len}}(C')}{\texttt{\emph{len}}(C_n)} \approx [2:1, n:1], \ \mathrm{where} \ n\geq3 \ ,
\end{equation}

\noindent and denotes a \emph{perfect entropy} if and only if a ratio of 1$:$1 after data decompression preserved, validating results for our data integrity. The use of $\texttt{\emph{len}}$ is to formally get the LOF (length-of-file) prior to other types of length functions, given in \S\,\ref{section2} and Theorem 2 in \S\,\ref{section5}, from the point of compressed data to the point of decompression. Symbol $C_{r}$ denotes the \emph{compression ratio} for all transmitted data from one data compression layer to another. To maintain the entropy from initial data, $C_{-1}$, thereafter, a data compression after four layers, and finally, another four layers conversely attaining data decompression, must be exact or indeed maintain a value of 1 (considering fuzzy quantum inclusions and exclusions, expectably). We could relate the previous relation to Eq.\,(\ref{4}), and thus to the following:
\begin{equation*}\label{16}
C_{r} =\frac{\mathrm{Initial \ Data \ Size}}{\mathrm{Compressed \ Size}} \times \frac{\mathrm{Compressed \ Size}}{\mathrm{Decompressed \ Size}} =\frac{\texttt{\emph{len}}(C_{-1})}{\texttt{\emph{len}}(C_n)}\times \frac{\texttt{\emph{len}}(C_n)}{\texttt{\emph{len}}(C')}=1 \, .
\end{equation*}

\noindent We emphasize that one must not confuse this representation of Uncompressed or Decompressed Size with Initial Size, where the former applies to a property of Restoring Data with a Decompressed Size or Final Retrieved Size, i.e., a property of Decompressed Data, apart from the terminologies used in conventional academic texts and scientific publications. The current emphasis is on the \emph{initial data} inputted against a progressive compression, subsequently, a progressive decompression, preserving a ratio of 1$:$1 in data content and entropic analysis from source-to-sink layers.

We have not studied the \emph{data-rate savings} relevant to the communication lines between $C$ and $C'$ layers on a DB application. This subject is quite essential to consider, covering the diagrammatic aspects of Fig.\,1, such that, quantum noise inclusions and exclusions are indeed objective to data compression amounts. Data-rate savings in the current FBAR algorithm has a noisy estimate for a limited string of characters input at every $t_{i}$ = 10 milliseconds.

Once the load of input increases, say, the FBAR thread's stack size as small as 0.5 kilobyte, despite of $t_{i...n}$ discrete equal time intervals, time $t$ decelerates stack elements' count due to memory overrun (stack overflow in sign of concurrent program loops) based on LIFO memory transactions (load) from the FBAR program. This is subject to memory allocation procedures, threads and multi-core optimization, quite resolutely tackled within .NET and C language applications, provided by their latest versions of MS-Visual Studio$^{\text{\textcircled{\tiny R}}}$ library packages. In the current FBAR application, the objective was to show the possibility of the FBAR concept and conversion layers from one state to another in a primitive state in VB\,6, which is quite experimental and relevant to the nature of the FBAR algorithm in a less elaborated version of code representation.

\section{Conclusion and future remarks}
\vspace{4pt}
\label{section6}
The basics of how future generation computers as quantum computers decipher message and encode with complexity, have been discussed and illustrated in this paper. We elaborated on how FBAR algorithm could assist in many ways, such as classical computers promoting the concept of fuzzy binary to fuzzy qubinary by projecting bits into subspaces of quantum type, achieving uncanny degrees of retrievable lossless embedded data into halved wavelength signals.

In conclusion, the project's aim is to conclude in major, the FQAR objective, as if the rise and fall of oscillations between absolute logic states of 1's and 0's reiterate for a binary sequence of equal values with respect to their symmetry, expressed in \S\S\,\ref{section2} and \ref{section3.4}. The powerful deductions on the FBAR objectives led to an AND/OR quantum theory, a provable theory and empirically correct in \emph{n}-dimensional physical systems and their applications. In the minor, the experimental results will speak for themselves within the context of our research proposal \S\,\ref{section3.4.1}. There are lots of open problems and next steps to be taken into the FBAR/FQAR project. One major step is to intensify the research via the cooperation between bi-bottom lattice sites with fqubit registers as an upgrade to current qubit registers by Zhang \emph{et al.}, Arimondo \emph{et al.} and Phillips~\cite{{36-Zhang et al.,{38-Arimondo et al.},{39-B. Phillips}}}. Therefore, the goal is to use the \emph{coding theory and cryptography} for the collaboration between different quantum components in quantum computers.

\noindent Interesting simulations have been made by organizations such as IBM like \emph{qcl quantum computer simulator} and similar algorithms testing quantum computation at large~\cite{46-IBM}. An attempt was made in describing our FQAR theory which requires similar interpreters to take action for a resourceful implementation relative to the limits of polarities, quantum noise inclusion and \emph{sub-bit} (or \emph{data dot}) addressing. In the implementation, the idea of using FBAR as FQAR for implementing fuzzy qubinary features, has not been researched by others and merely discussed in discrete, vaguely finding connection between their domain of concepts. Here, the prime goal is to analyze the required overhead of memory systems in their organizations as well as introducing the service oriented architecture in detail, in order to analyze the novel features emerging out of the algorithm. Moreover, to extend the library of codes for encoded data, one implements protection for security engineering as well as novel computation of quantum systems between nodes and databases. With the extensions, FBAR could also be tested against intrusions prior to AND and OR operation, reaching a level of special decryption devices impossible to hack into systems with such a complex fuzzy qubinary computational ability.

As it was described, the data compression is highly ranked once promoted to FQAR and the theory, hereby laid out the standards to its implementation. The FBAR current state could supply networks with a factor of 2$\sim$3:1 compression and ranked amongst the strongest compressors across the globe. Imagining the $2n$$:$$1$ compression ratio is not far fetched and the theory of its information has been discussed in this report.

We aim to achieve such standards for future generation computers, concluding our research as a first-grade accomplishment outsmarting present technologies in their ability of computation, which is too variant and barely scratch the surface in terms of fractal allocation models on data with the same integrity per se. This research benefits a class of combinatorics between ``data dot" interactions, more likely to move out from theoretic-based applications and research labs into practical applications. Finally, treating data types with this FBAR AND/OR data compression technique promotes optimum online media technology with significant lossless data compression outcomes.

\section*{Acknowledgments}
\vspace{4pt}
The work described in this paper was encouraged by C. Svahnberg, reviewed by M. Boldt and S. Axelsson from the Faculty \textit{of }Department \textit{of} Software Engineering and Computer Science, Blekinge Institute \textit{of }Technology, Sweden (2009). Their constructive remarks led me to maintain coherence in representing the concept, thereby choosing relevant references assimilated within this report.

\end{document}